\documentstyle[pra,aps]{revtex}
\draft
\begin{document}

\author{J. P. Clemens$^*$, P. R. Rice$^*$, and L. M. Pedrotti$^+$}

\address{$^*$Department of Physics, Miami University, Oxford, OH  45056\\
$^+$Department of Physics, University of Dayton, Dayton, OH 45469}
\title{Output Spectrum of Single-Atom Lasers}
\date{\today }
\maketitle

\begin{abstract} We consider a laser composed of a single atom in a 
microcavity, with a coherent or incoherent pump. We consider both three- and 
four-level gain schemes, and examine the output spectrum of such lasers. We 
find that the linewidth generally scales as the inverse of the photon number. 
For large atom-field coupling, a vacuum-Rabi doublet structure is obtained. In 
the three-level case, this vacuum-Rabi splitting is apparent only for small 
intracavity photon numbers, and vanishes for large pumps. In the four-level 
scheme, the vacuum-Rabi structure appears at a nonzero pump level, and is 
maintained for large pumps, even when the intracavity photon number is larger 
than unity. This behavior is explained utilizing the quantum trajectory 
approach.
\end{abstract}

\section{Introduction}
In recent years it has become possible to explore the dynamics of a single 
atom as it passes through a microcavity at low velocity. By this we mean that 
the transit time for the atom across the cavity mode is on the order of a 
hundred spontaneous emission lifetimes. This slow atomic beam is generated by 
dumping cold atoms from a magneto-optical trap into a microcavity 
\cite{Kimble}. There has been much work in the past on the 
micromaser/microlaser, where two-level atoms in the excited state are flown 
through a microcavity of high-Q, and interact with the field mode of the 
cavity \cite{micromaser}.  These systems are quite interesting and exhibit a 
variety of nonclassical effects, but we wish to examine here the case where 
the atom starts in the ground state, lives in the cavity mode for many 
lifetimes with an essentially constant atom-field coupling strength. Of course 
this coupling strength will change with time as the atom moves through the 
Gaussian profile of the mode, but we will assume here that the atom-field 
dynamics are such that the atom is essentially stationary. Hence we have the 
one-atom limit of a laser pointer; we have a fixed gain medium in a cavity, 
pumped by an external source, but the gain medium is composed of a single 
three- or four-level atom. Some previous work has been done on this system.
Smith and Gardiner \cite{gardiner1} were the first to consider such a system, 
but did not explore large enough atom-field couplings to obtain interesting 
results. The key result of their paper was a way to treat single atom systems 
in a Fokker-Planck approach.  Mu and Savage \cite{MuSavage} considered several 
classes of single atom lasers and focused on the photon statistics. They 
showed that lasing was possible for such a system, and that for large 
atom-field coupling the single atom laser emitted amplitude squeezed, or 
antibunched light. For parameters for which the output light was amplitude 
squeezed, the linewidth of the laser was shown to increase with pump strength 
rather than decrease as in the usual Schawlow-Townes type fashion.  Ritsch and 
Pellazari \cite{RitschPell} also examined the photon statistics of a single 
atom laser, and they too predict regions of parameter space where the output 
is amplitude squeezed.  Ginzel et. al. \cite{Ginzel} considered various 
aspects of single-atom laser systems, including the observation of a 
vacuum-Rabi doublet in the output spectrum, but in their work the emphasis was 
on the application of a new computational approach, that of the damping basis. 
In that approach, the basis states were eigenstates of the dissipative part of 
the master equation.  Loffler et. al. \cite{Loffler} considered the spectrum 
of a three-level single atom laser, and predicted vacuum-Rabi structures in 
the output spectrum. In this work we show that the two-peaked vacuum-Rabi 
structure vanishes quickly as the pump strength is increased. More recently 
Jones et. al. \cite{Jones} have examined the photon statistics of single-atom 
laser systems, with particular emphasis on how the systems behaved as $\beta$ 
was changed. The parameter $\beta$ is the fraction of spontaneous emission 
into the lasing cavity. It is a parameter of much interest in the microlaser 
community. As $\beta$ tends towards unity, the laser output rises linearly 
with pump strength; this has been referred to as a ``thresholdless" laser. For 
more on this subject, we refer the reader to recent work on macroscopic laser 
systems and their dependence on $\beta$. \cite{HJCPRR,Yamamoto}

In section 2, we discuss the output spectrum of a three-level incoherently 
pumped single-atom laser. Section 3 deals with the four-level incoherently 
pumped single-atom laser. In section 4 we examine the four-level model, but 
with coherent pumping, to determine if changing the pumping mechanism alters 
the results. Finally in section 5 we conclude.

A schematic of the system is shown in Figure \ref{fig:sal1}. We adiabatically 
eliminate the fast transition from the topmost state to the upper lasing 
level, and use a two-level model, where the incoherent pump is modeled via the 
$\Gamma$ term in the following master equation, \cite{Haken}
\begin{eqnarray}
\dot \rho &=&{{-i} \over \hbar }[H_s,\rho ]+\kappa (2a\rho 
\,a^\dagger-a^\dagger a\rho -\rho \,a^\dagger )\nonumber \\
&&+{{\gamma _{}} \over 2}(2\sigma _-\rho \sigma _+-\sigma _+\sigma _-\rho 
-\rho \sigma _+\sigma _-)\nonumber \\
&&+{\Gamma  \over 2}(2\sigma _+\rho \sigma _--\sigma _-\sigma _+\rho -\rho 
\sigma _-\sigma _+)
\end{eqnarray}
with the system Hamiltonian given by
\begin{equation}
H_S=i\hbar g(a^\dagger \sigma _--a\sigma _+)
\end{equation}

The only nonzero density matrix elements are the diagonal elements 
(populations of the various levels) and coherences between states $\mid n,1 
\rangle$ and $\mid n-1, 2 \rangle$. Here $n$ denotes the photon occupation 
number. The equations for these density matrix elements are
\begin{eqnarray}
{{d\rho _{n,1;n,1}} \over {dt}}&=&2 \kappa (n+1)\rho _{n+1,1;n+1,1} - 
\left( 2 \kappa  n+\Gamma \right) \rho _{n,1;n,1}\nonumber \\
&&+\gamma \rho _{n,2;n,2}-2\sqrt {n+1}\kern 1pt g\rho _{n,1;n-1,2}\\
  {{d\rho _{n,2;n,2}} \over {dt}}&=&2\kappa (n+1) \rho _{n+1,2;n+1,2} 
  -\left\{  2 \kappa  n+\gamma  \right\}\rho _{n,2;n,2}\nonumber \\
&&+\Gamma \rho _{n,1;n,1}+2\sqrt n\kern 1pt g\rho _{n,1;n-1,2}\\
  {{d\rho _{n,1;n-1,2}} \over {dt}}&=&2\kern 1pt \kappa \kern 1pt \sqrt 
  {n(n-1)}\kern 1pt \rho _{n+1,2;n,1}
 \nonumber \\ &&-\left\{ {\kappa (2n\kern 1pt -1)
+{{\Gamma +\gamma} \over 2} } \right\} \rho _{n,2;n-1,1}\nonumber \\
&&+\sqrt n g\left( {\rho _{n-1,2;n-1,2}-\rho _{n,1;n,1}} \right)
\end{eqnarray}
We solve these equations in the steady state, to obtain needed initial 
conditions for spectrum calculations. These matrix elements can also be used 
to calculate photon statistics, such as the mean photon number and Fano factor 
(variance over the mean). Typically we start by truncating the photon basis at 
some small number (3-10). Calculations are checked at the end to check that 
the population of the highest photon number states is less than $10^{-4}$. If 
that were not the case, the program that does our calculations repeats the 
process, but increments the maximal photon number, and rechecks that we are 
keeping enough photon states. The calculation uses $\sum\limits_{i,n} 
\rho_{i,n;i,n}=1$ to solve for $\rho_{0,-;0,-}$ in terms of the other diagonal 
density matrix elements.  In much of what follows, we will refer to $\beta$, 
the fraction of spontaneous emission into the cavity mode.
\begin{equation}
\beta ={{2g^2/(\gamma +\Gamma +2\kappa )} \over {2g^2/(\gamma +\Gamma 
+2\kappa )\;+\gamma /2}}
\end{equation}

We are interested in calculating the output spectrum of the laser, 
\begin{equation}
S(\omega ) = \int\limits_{-\infty }^\infty  {d\tau \,e^{i\omega 
\,\tau }}\left\langle {a^\dagger(0)\,a\,(\tau )} \right\rangle =2\Re 
\int\limits_0^\infty  {d\tau \,e^{i\omega \,\tau }}\left\langle 
{a^\dagger(0)\,a\,(\tau )} \right\rangle.
\end{equation}
We calculate this spectrum using the quantum regression theorem, 
\begin{equation}
\left\langle {a^+(0)\,a\,(\tau )} \right\rangle =tr\left\{ {a(0)\,A(\tau )} 
\right\}=\sum\limits_{i,n} {\sqrt {n+1}\,\left\langle {i,n+1\,|\,A(\tau ) 
\,|i,n} \right\rangle }
\end{equation} 
where 
  $A(0)=\rho _{SS}\,a^\dagger$ and $ \dot A={\cal L}A$. 
The resulting equations can be written in the form
\begin{equation}
{{d\vec A } \over {dt}}=\mathord{\buildrel{\lower3pt\hbox{$ 
\scriptscriptstyle\leftrightarrow$}}\over M}\vec A
\end{equation}

The relevant equations for the matrix elements of $\vec A$ are
\begin{eqnarray}
{{dA_ {n+1,1;n,1}} \over {dt}}&=&2\kappa \sqrt{(n+2)(n+1)}A_{n+2,1;n+1,1} 
\nonumber
\\ &&
-\left( \kappa (2n+1)+\Gamma \right) A_{n+1,1;n,1}+\gamma A_{n+1,2;n,2}
\nonumber \\&&
+g\sqrt{n+1}A_{n,2;n,1}+g\sqrt{n}A_{n+1,1;n-1,2}\\
{{dA_ {n+1,2;n,2}} \over {dt}}&=&2\kappa\sqrt{(n+2)(n+1)}A_{n+2,2;n+1,1}
\nonumber \\&&-\left( \kappa (2n+1) +\gamma \right) A_{n+1,2;n,2}+\Gamma 
A_{n+1,1;n,1}
\nonumber \\ &&-g\sqrt{n+2}A_{n+2,1;n,2} -g\sqrt{n+1}A_{n+1,2,n+1,1}\\
{{dA_ {n+2,1;n,2}} \over {dt}}&=&2\kappa \sqrt{(n+3)(n+1)}A_{n+3,1;n+1,2}
\nonumber \\
&&
-\left( \kappa (2n+1)+\gamma /2 +\Gamma /2 \right) A_{n+2,1;1,2}
\nonumber \\&&+g\sqrt{n+2}A_{n+1,2;n,2}-g\sqrt{n+1}A_{n+2,1;n+1,1}\\
{{dA _{n,2;n,1}} \over {dt}}&=&2\kappa (n+1)A_{n+1,2;n,1} -\left(
2\kappa +\gamma /2+\kappa /2 \right)\nonumber \\
&&+g\sqrt{n}A_{n,2;n-1,2}-g\sqrt{n+1}A_{n+1,1;n,1}
\end{eqnarray}
After taking the Fourier transform of the above differential equations we have
\begin{equation}
 \vec {\tilde A}(\omega )={\left\{ \mathord{\buildrel{\lower3pt\hbox{$ 
 \scriptscriptstyle\leftrightarrow$}}\over M}-i\omega
\mathord{\buildrel{\lower3pt\hbox{$\scriptscriptstyle\leftrightarrow$}} 
\over I} \right\}}^{-1} \vec A (0)
\end{equation}
with $\vec {\tilde A}(\omega )$ composed of the Fourier transform of $\vec A 
(\tau)$ and then we can easily form the spectrum
\begin{equation}
S(\omega )=\sum\limits_{i,n} {\sqrt {n+1}\,\left\langle {i,n+1\,|\,\Re 
\;\tilde A(\omega )\,|i,n} \right\rangle }.
\end{equation}
In solving these equations, we truncate the photon basis at the same value of 
$n$ as in the density matrix element equations.

In Figure \ref{fig:sal2}, we plot the output spectrum of the laser for 
$g/\gamma=0.1$ and $\kappa /\gamma=0.1$ as a function of pumping strength, 
$\Gamma$. We see that the lineshape is approximately Lorentzian, and the width 
decreases initially as the pump strength is increased. All spectra in this 
paper are normalized so that the integrated spectrum is unity. At larger pumps 
we see that the linewidth begins to broaden. This is apparent most readily 
from examining the peak of the spectrum. It rises rapidly with pump rate, 
reaches a maximum, and slowly goes back down. As the area is the same (by our 
normalization) this behavior is indicative of the initial narrowing, which 
reaches a minimum before beginning to broaden. In Figure \ref{fig:sal3}, we 
plot the linewidth, obtained by fitting a Lorentzian curve, versus pump 
strength.  For comparison, we also plot $\Delta \omega_{ST}=\kappa /2\langle 
n \rangle$, the Schawlow-Townes result. Here the mean photon number $\langle 
n \rangle$ has been directly calculated from the steady state density matrix 
elements via
$\langle n \rangle = \sum\limits_{n,j} n \langle n,j \mid \rho_{SS} \mid n,j 
\rangle $
Here $n$ of course is the photon number index, and $j=1,2$ are the atomic 
level indices. We see that while the linewidth initially decreases with photon 
number, it is always broader than the Schawlow-Townes result for small pump 
rates. We see that as the laser turns off with increasing pump strength, the 
linewidth increases, and goes below the Schawlow-Townes result, but 
qualitatively it follows that trend.  The mean intracavity photon number is 
plotted as a function of driving field strength, for these same parameters, 
in Figure \ref{fig:sal4}. We see that the laser does turn off with pump 
strength. This is a feature of the incoherently pumped three-level laser, both 
for single atom and macroscopic systems, as shown by Mu and Savage 
\cite{MuSavage}, Jones et. al. \cite{Jones}, and Koganov and Shuker 
\cite{Koganov}. This is due to the incoherent nature of the pumping process, 
which causes the atom to uncouple from the field for high pump rates. This is 
due to the incoherent pump process decohering the induced dipole on the lasing 
transition. This can also be viewed as $\beta \rightarrow 0$ as $\Gamma 
\rightarrow \infty$, that the fraction of spontaneous emission into the cavity 
mode is pump dependent, and there is no spontaneous emission into the cavity 
for high pump rates. From the usual arguments of quantum electrodynamics, 
there is also no stimulated emission. From the viewpoint of quantum trajectory 
theory,  the pump mechanism is a jump process, and the atom becomes trapped in 
the upper state of the lasing transition. As one increases $g/\gamma$ to 0.6, 
we have found that the single atom laser emits amplitude squeezed light 
\cite{Jones}, and that the linewidth indeed increases with pump strength, even 
for small pumps, as first predicted by Mu and Savage \cite{MuSavage}. Further 
increasing $g/\gamma$ to $1.414$, as in Figure \ref{fig:sal5}, we see that the 
spectrum exhibits a double-peaked structure. This has been predicted by 
Loffler et. al. \cite{Loffler}, and they identified the source of this 
structure vacuum-Rabi oscillations on the lasing transition. However we see 
that this structure only exists for very small pump strengths, when there are 
not very many photons in the cavity. This is not unexpected, as the 
vacuum-Rabi oscillations are associated with coherent oscillations of the 
one-photon states $\mid 0,2 \rangle $ and $\mid 1,1 \rangle $. When states 
with higher photon number are occupied, the vacuum-Rabi structure vanishes, as 
many transition frequencies between various dressed states start to appear.  
In Figure \ref{fig:sal6}, we show the  spectrum for larger pump values, where 
one can observe the vacuum-Rabi doublet disappear. Essentially the doublet is 
coming from spontaneous emission of the strongly coupled atom-cavity system, 
and in no sense from a laser. The mean photon number versus pump is exhibited 
in Figure \ref{fig:sal7}. There is of course no well defined threshold for 
such a microscopic case. We see that the vacuum-Rabi oscillations vanish well 
before the mean photon number nears unity. Returning our attention to Figure 
\ref{fig:sal6} for a moment, we notice that again in this case, at high pump 
rates when the mean intracavity photon number begins to decrease, the 
linewidth begins to increase. Again this is easily apparent from the drop in 
the value of the normalized spectrum at line center, and is consistent with 
the turn off of the laser.

We gain further insight into this behavior by examining results of quantum 
trajectory simulations. \cite{Howardbig,Dalibard} The conditioned wave 
function is taken to be 
\begin{equation}
|\psi _c(t)\rangle =\sum\limits_{n=0}^\infty  {}C_{1,n}(t)e^{-iE_{1,n}t}| 
1,n\rangle +C_{2,n}(t)e^{-iE_{2,n}t}|2,n\rangle \end{equation}
The coherent evolution of the conditioned wave function obeys the Schroedinger 
equation with the following non-Hermitian Hamiltonian,
\begin{eqnarray}
H_D&=&\hbar (\omega -i\kappa )a^+a+i\hbar g\left( a^+\sigma _{32}- 
a\sigma _{-}\right) \nonumber \\
 &&-i\hbar {\gamma  \over 2}\sigma _{+}\sigma _{-}-i\hbar {\Gamma  
 \over 2}\sigma _{-}\sigma _{+}
\end{eqnarray}
At each time step, the system is subject to collapses of the wavefunction 
according to the collapse operators
\begin{eqnarray}
\hat F_1&=&\sqrt \gamma \,\sigma _{-}\\
\hat F_2&=&\sqrt \Gamma \,\sigma _{+}\\
\hat F_3&=&\sqrt {2\kappa }\,a.
\end{eqnarray}
The probability of a collapse is proportional to the size of the time step 
multiplied by $\langle \psi_{c} \mid F^\dagger F \mid \psi_{c}\rangle$. A 
separate random number is used to determine at each time step whether a 
particular collapse occurs. In the unlikely case that two collapses are to 
occur in a given time step, another random number is used to determine which 
actually occurs. Of course, we must choose our time step to be small compared 
to the fastest rates in the problem to minimize such occurrences. In Figures 
\ref{fig:sal8}-\ref{fig:sal9} we plot the induced dipole for the lasing 
transition, and population of the upper lasing state for two values of 
$\Gamma /\gamma$ and $g/\gamma=1.414$. For smaller pump rates, as in Figure 
\ref{fig:sal8}, there is an obvious vacuum-Rabi oscillation apparent. For 
larger pump strengths, as in Figure \ref{fig:sal9},  there are effectively no 
coherent oscillations in the induced dipole or population, hence there is no 
vacuum-Rabi structure for large pumps. The incoherent pump process interrupts 
the coherent oscillations. In the quantum trajectory view, the incoherent pump 
is modeled as an upward jump. A pump event places the atom in the upper state 
of the lasing transition, and kills off the coherence between the two lasing 
states. As the pump rate increases, the atom becomes trapped in the upper 
level of the lasing transition, and decouples from the field. This eventually 
results in the mean photon number dropping to zero as the pump rate is 
increased.

\section{Incoherently Pumped Four-Level Laser}
\label{sec:sal3}
This system is shown schematically in Figure \ref{fig:sal10}. Again, we 
adiabatically eliminate the level above the upper lasing level. We are left 
with an effective three-level system, with the incoherent pump modeled as in 
the above work on the incoherently pumped three-level laser. The master 
equation for the incoherently pumped four-level laser is then
\begin{eqnarray}
\dot \rho&=&{{-i} \over \hbar }[H_s,\rho ]+\kappa (2a\rho \,a^\dagger 
-a^\dagger a\rho -\rho \,a^\dagger a)\nonumber\\
&&+{\Gamma  \over 2}(2\sigma _{13}\rho \sigma _{31}-\sigma _{31} 
\sigma _{13}\rho -\rho \sigma _{31}\sigma _{13})\nonumber\\
&&+{\gamma  \over 2}(2\sigma _{32}\rho \sigma _{23}-\sigma _{23} 
\sigma _{32}\rho -\rho \sigma _{23}\sigma _{32})\nonumber \\
&&+{{\gamma _f} \over 2}(2\sigma _{21}\rho \sigma _{12}-\sigma _{12} 
\sigma _{21}\rho -\rho \sigma _{12}\sigma _{21})
\end{eqnarray}
with
\begin{equation}
H_S=i\hbar g(a^\dagger \sigma _{32}-a\sigma _{23})
\end{equation}
and the following definition for atomic raising and lowering operators
\begin{equation}
\sigma _{ij}=\,|\kern 1pt j\rangle \,\langle i\,|.
\end{equation}

The equations for the nonzero density matrix elements are
\begin{eqnarray}
{{d\rho _{n,1;n,1}} \over {dt}}&=&2\kappa (n+1) \rho _{n+1,1;n+1,1} 
-\left\{  2 \kappa  n+\Gamma  \right\} \rho _{n,1;n,1}\nonumber \\
&&+\gamma _f \rho _{n,2;n,2} \\
{{d\rho_{n,2;n,2}} \over {dt}}&=&2 \kappa  (n+1) \rho _{n+1,2;n+1,2}- 
\left\{ 2 \kappa n+\gamma _f \right\} \rho _{n,2;n,2}\nonumber \\
&&+\gamma\rho _{n,3;n,3}+2\sqrt n g\rho _{n,2;n-1,3}\\
 {{d\rho _{n,3;n,3}} \over {dt}}&=&2 \kappa  (n+1) \rho_{n+1,3;n+1,3}- 
 \left\{  2 \kappa  n+\gamma  \right\}  \rho_{n,3;n,3}\nonumber \\
&&+\Gamma \rho _{n,1;n,1}-2\sqrt {n+1}\kern 1pt g\rho _{n,3;n+1,2}\\
 {{d\rho _{n,2;n-1,3}} \over {dt}}&=&2 \kappa \sqrt {n(n-1)}\rho _{n+1,2;n,3}
 -\left\{ {\kappa (2n -1)+{\gamma  \over 2}} \right\}\rho _{n,2;n-1,3}
 \nonumber \\
&&+\sqrt n g\left\{ {\rho _{n-1,3;n-1,3}-\rho _{n,2;n,2}} \right\}.
\end{eqnarray}
We again calculate this spectrum using the quantum regression theorem.
The relevant equations for the matrix elements of $A$ in this case are
\begin{eqnarray}
{{dA_ {n+1,1;n,1}} \over {dt}}&=&2\kappa \sqrt{(n+2)(n+1)}A_{n+2,1;n+1,1}
-\left( \kappa (2n+1) +\Gamma \right)A_{n=1,1;n,1}\nonumber \\ &&+\gamma_f 
A_{n+1,2;n,2},\\
{{dA _{n+1,2;n,2}} \over {dt}}&=&2\kappa \sqrt{(n+2)(n+1)}A_{n+2,2;n+1,2}
-\left( \kappa (2n+1)+\gamma_f \right) A_{n+1,2;n,2} \nonumber \\  &&+\gamma 
A_{n+1,3;n,3}
+g\sqrt{n+1}A_{n,3;n,2}+g\sqrt{n}A_{n+1,2;n-1,3},\\
{{dA_ {n+3,1;n,3}} \over {dt}}&=&2\kappa \sqrt{(n+2)(n+1)}A_{n+2,3;n+1,3}
-\left( \kappa (2n+1)+\gamma \right)A_{n+1,3;n,3}\nonumber \\ &&+\Gamma 
A_{n+1,1;n,1}
-g\sqrt{n+2}A_{n+2,2;n,3}-g\sqrt{n+1}A_{n+1,3;n+1,2},\\
{{dA _{n+2,2;n,3}} \over {dt}}&=&2\kappa \sqrt{(n+3)(n+1)}A_{n+3,2;n+1,3}
\nonumber \\
&&-\left( \kappa (2n+2) +\gamma /2 +\gamma_f /2 \right) A {n+2,2;n,3}
\nonumber \\ &&+g\sqrt{n+2}A_{n+1,3;n,3}-g\sqrt{n+1}A_{n+2,2;n+1,2},\\
{{dA _{n,3;n,2}} \over {dt}} &=&2\kappa (n+1)A_{n+1,3;n+1,2}-\left(
2\kappa n +\gamma /2 \gamma_f /2 \right) A_{n,3;n,2}\nonumber \\
&&+g\sqrt{n}A_{n,3;n-1,3}-g\sqrt{n+1}A_{n+1,2;n,3}.
\end{eqnarray}
After taking the Fourier transform of the above equations, we have
\begin{equation}
 \vec {\tilde A}(\omega )={\left\{ \mathord{\buildrel{\lower3pt\hbox{$ 
 \scriptscriptstyle\leftrightarrow$}}\over M}-i\omega
\mathord{\buildrel{\lower3pt\hbox{$\scriptscriptstyle\leftrightarrow$}}
\over I} \right\}}^{-1} \vec A (0)
\end{equation}
with $\vec {\tilde A}(\omega )$ composed of the Fourier transform of $
\vec A (\tau)$ and then we can easily form the spectrum
\begin{equation}
S(\omega )=\sum\limits_{i,n} {\sqrt {n+1}\,\left\langle {i,n+1\,|\,\Re 
\;\tilde A(\omega )\,|i,n} \right\rangle }
\end{equation}
In solving these equations, we truncate the photon basis at the same value of 
$n$ as in the density matrix element equations.

In Figure \ref{fig:sal11}, we plot the spectrum of the laser for 
$g/\gamma=1.4, \gamma_f /\gamma=10.0$ and $\kappa /\gamma =0.1$. We see that 
the spectrum is a single peaked structure, whose linewidth decreases with pump 
strength initially, and asymptotically approaching a limiting value. In Figure 
\ref{fig:sal12}, we plot the linewith of the laser spectrum versus pump 
strength, for various values of $\beta $, for $\kappa /\gamma=0.1$ and 
$\gamma_f /\gamma=10.0$. Recall that the fraction of spontaneous emission into 
the lasing cavity, $\beta$, is then determined by $g$, with $\kappa$, 
$\gamma_f$, and $\gamma$ fixed. This linewidth is again obtained by curve 
fitting a Lorentzian to the output spectrum. For comparison, we also plot the 
Schawlow-Townes result, $\kappa /2 \langle n \rangle$; the mean photon number 
again calculated from the steady state density matrix. We see that the 
linewidth is always broader than the Schawlow-Townes limit, but that the 
linewidth decreases with the inverse of the photon number. The photon number 
pins for large pump rates, as it does no good to pump the single atom laser 
faster than the fastest decay rate (usually $\gamma_f$), and the linewidth 
also pins at an asymptotic value. As one increases $\beta  $ to the range 
$0.6 - 0.9$, (or increases $g$ given that $\kappa /\gamma$ and $\gamma_f 
/\gamma$ are fixed), the asymptotic value of the linewidth for large pumps 
begins to increase, as shown in Figure \ref{fig:sal13}.  For a value of 
$\beta=0.998$, with $\kappa /\gamma=0.1$, and $\gamma_f /\gamma=100.0$, we see 
that the linewidth increases with pump strength. As in the case of the 
three-level system, this is concurrent with the emitted light being amplitude 
squeezed. In fact for all values of $\beta$ above $0.5$ or so, the light is 
amplitude squeezed \cite{Jones}. For the uppermost curve in Figure 
\ref{fig:sal13}, $g/\gamma=100.0$, but there is no chance of vacuum-Rabi 
oscillations as $\gamma_f /\gamma=100.0$, and so the decoherence rate 
$\gamma_f=g$. We can obtain vaccum-Rabi structure in the output spectrum in 
many cases, for example $g/\gamma=10.0$, $\gamma_f /\gamma=2.0$, and $\kappa 
/\gamma=0.1$ as shown in Figure \ref{fig:sal14}.  We see that there is a 
single peaked structure at small pump rate, which then evolves into a double 
peaked, vacuum-Rabi structure at large pumps, reaching an asymptotic spectrum 
for large pumps. The average photon number in the cavity reaches $\langle n 
\rangle=2.6$  for those parameters and large pump, and is larger than one, as 
shown in Figure \ref{fig:sal15}. The location of the two peaks are not well 
approximated by the complex part of the single-photon eigenvalues for this 
system, nor is the width well approximated by the real part. This is due to 
the fact that more than the one-photon state is involved in this process.

To understand this behavior, it is again instructive to look at quantum 
trajectory simulations. We take the conditioned wave function to be
\begin{equation}
|\psi _c(t)\rangle =\sum\limits_{n=0}^\infty  {}C_{1,n}(t)e^{-iE_{1,n}t}
|1,n\rangle +C_{2,n}(t)e^{-iE_{2,n}t}|2,n\rangle +C_{3,n}(t)e^{-iE_{3,n}t}
|3,n\rangle 
\end{equation}
where again, the unitary evolution of this wave function is governed by a 
Schrodinger equation with the following non-Hermitian Hamiltonian,
\begin{eqnarray}
H_D&=&\hbar (\omega -i\kappa )a^+a+i\hbar g(a^\dagger\sigma _{32}-a
\sigma _{23})\nonumber \\
  && -i\hbar {\gamma  \over 2}\sigma _{23}\sigma _{32}-i\hbar {{\gamma _f} 
  \over 2}\sigma _{12}\sigma _{21}-i\hbar {\Gamma  \over 2}\sigma _{31}
  \sigma _{13}.
\end{eqnarray}
Here we have four associated collapse processes, that are governed by the 
following four collapse operators,
\begin{eqnarray}
\hat F_1&=&\sqrt \gamma \,\sigma _{32}\\
\hat F_2&=&\sqrt {\gamma _f}\,\,\sigma _{21}\\
\hat F_3&=&\sqrt \Gamma \,\sigma _{13}\\
\hat F_4&=&\sqrt {2\kappa }\,a.
\end{eqnarray}
These trajectories are generated in the same manner as those of the 
three-level system in the proceeding section. The derivation of $\beta$ is 
particularly transparent in the quantum trajectory formalism, using the 
equations for the probability amplitudes for the various states,
\begin{eqnarray}
\dot C_{1,n}&=&-\left( {{\Gamma  \over 2}+n\kappa } \right)C_{1,n}\\
\dot C_{2,n+1}&=&-\left( {{{\gamma _f} \over 2}+(n+1)\kappa } \right)
C_{2,n+1}+g\sqrt {n+1}C_{3,n}\\
\dot C_{3,n}&=&-\left( {{{\gamma _{}} \over 2}+n\kappa } \right)
C_{3,n}-g\sqrt {n+1}C_{2,n+1}.
\end{eqnarray}

If $\gamma _f>>\gamma ,g,\kappa ,\Gamma $, then we have
\begin{equation}
\dot C_{3,n}=-\left( {{\gamma  \over 2}+n\kappa } \right)C_{3,n}-{{g^2} 
\over {\kappa (n+1)+\gamma _f/2}}(n+1)C_{1,n+1}
\end{equation}
In the case of $n=0$, we may read off
\begin{equation}
\beta ={{2g^2/(\gamma _f+2\kappa )} \over {2g^2/(\gamma _f+2\kappa )\;
+\gamma /2}}
\end{equation}
as the fraction of spontaneous emission into the cavity mode.

Why is there no double-peaked structure in the spectrum for small pumps? 
Examining the temporal evolution of the induced dipole on the lasing 
transition in Figure \ref{fig:sal16}, we see that for small pump strengths 
the dipole is usually zero, and an essentially random time occurs before the 
next oscillation occurs, which then lasts for some variable length of time.  
We see similar behavior in the population of the upper lasing level.  For 
larger pump strengths, as in Figure \ref{fig:sal17}, the dipole is often 
interrupted by a jump to the ground state of the system (a $\gamma_f$ event), 
which is then swiftly followed by a pump event. This sequence most often 
occurs when the atom has vacuum-Rabi flopped to the lower lasing level. Then 
the coherent vacuum-Rabi oscillations are begun again. The interruptions due 
to the collapses also broaden the vacuum-Rabi peaks, but the dipole is still 
mainly periodic if not pure sinusoidal.  Again, we see similar behavior in the 
population of the upper lasing level.  (We also note that the two peaked 
structure remains even when the mean intracavity photon number is well above 
one.)

\section{Four-Level Coherently Pumped Laser}
\label{sec:sal4}
\ \indent In this section we examine a coherently pumped four-level single 
atom laser, which is shown schematically in Figure \ref{fig:sal18}. The master 
equation for this system is given by
\begin{eqnarray}
\dot \rho &=&{{-i} \over \hbar }[H_s,\rho ]+\kappa (2a\rho \,a^\dagger 
-a^\dagger a\rho -\rho \,a^\dagger a)\nonumber \\
&&+{\gamma  \over 2}(2\sigma _{32}\rho \sigma _{23}-\sigma _{23}
\sigma _{32}\rho -\rho \sigma _{23}\sigma _{32})\nonumber \\
&&+{{\gamma _f} \over 2}(2\sigma _{21}\rho \sigma _{12}-\sigma _{12}
\sigma _{21}\rho -\rho \sigma _{12}\sigma _{21}),
\end{eqnarray}
with
\begin{equation}
H_S=i\hbar g(a^\dagger \sigma _{32}-a\sigma _{23})+i\hbar E_{pump}(
\sigma _{41}-\sigma _{14}),
\end{equation}
where again $\sigma _{ij}=\,|\kern 1pt j\rangle \,\langle i\,|$.

The nonzero density matrix elements satisfy the following equations,
\begin{eqnarray}
{{d\rho _{n,1;n,1}} \over {dt}}&=&2 \kappa  (n+1) \rho _{n+1,1;n+1,1}
\nonumber \\
&&-2\kappa n \rho _{n,1;n,1}
+\gamma _{21} \rho _{n,2;n,2}+2\Gamma \rho _{n,4;n-1,1}\\
 {{d\rho _{n,2;n,2}} \over {dt}}&=&2\kappa  (n+1) \rho _{n+1,2;n+1,2}- 
 \left\{ { \kappa n
+\gamma _{21}} \right\}\rho _{n,2;n,2}\nonumber \\
&&+\gamma _{}\rho _{n,3;n,3}+2\sqrt n\kern 1pt g\rho _{n,2;n-1,3}\\
 {{d\rho _{n,3;n,3}} \over {dt}}&=&2\kern 1pt \kappa \kern 1pt (n+1)\kern 
 1pt \rho _{n+1,3;n+1,3}-\kern 1pt \,\kern 1pt \left\{ \kappa \kern 1pt \kern 
 1pt n 
+\gamma  \right\} \rho _{n,3;n,3} \nonumber \\
&&+\gamma _{43}\rho _{n,1;n,1}-2\sqrt {n+1}\kern 1pt g\rho _{n,3;n+1,2}\\
  {{d\rho _{n,4;n,4}} \over {dt}}&=&2\kern 1pt \kappa \kern 1pt (n+1)\kern 
  1pt \rho _{n+1,3;n+1,3} \nonumber \\
&&-\kern 1pt \,\kern 1pt \left\{ {\kern 1pt 2\kern 1pt \kappa \kern 1pt 
\kern 1pt n
+\gamma _{43}} \right\}\rho _{n,3;n,3}-2\Gamma \rho _{n,4;n-1,1}\\
  {{d\rho _{n,2;n-1,3}} \over {dt}}&=&2\kern 1pt \kappa \kern 1pt \sqrt 
  {n(n-1)}\kern 1pt \rho _{n+1,2;n,3}-\kern 1pt \,\kern 1pt \left\{ {\kappa 
  (2n\kern 1pt -1)+{\gamma  \over 2}} \right\}\rho _{n,2;n-1,3}\nonumber \\
&&+\sqrt n\kern 1pt g\left\{ {\rho _{n-1,3;n-1,3}-\rho _{n,2;n,2}} \right\}\\
  {{d\rho _{n,1;n,4}} \over {dt}}&=&2\kern 1pt \kappa \kern 1pt \sqrt 
  {n(n-1)}\kern 1pt \rho _{n+1,1;n+1,4}-\kern 1pt \,\kern 1pt \left\{ {\kappa 
  (2n\kern 1pt -1)+{\gamma  \over 2}} \right\}\rho _{n,1;n,4}\nonumber \\
&&+\Gamma \left\{ {\rho _{n,4;n,4}-\rho _{n,1;n,1}} \right\}.
\end{eqnarray}

We again calculate this spectrum using the quantum regression theorem.
The relevant equations for the matrix elements of $\vec A$ in this case are
\begin{eqnarray}
{{dA _{n+1,1;n,1}} \over {dt}}&=&2\kappa \sqrt{(n+2)(n+1)}A_{n+2,1;n+1,1}
-\left( \kappa (2n+1)\right)A_{n+1,1;n,1}+\gamma_f A_{n+1,2;n,2}+\nonumber \\  
&& E \left( C_{n+1,4;n,1}+C_{n+1,1;n,4} \right)
\\
{{dA_ {n+1,2;n,2}} \over {dt}}&=&2\kappa \sqrt{(n+2)(n+1)}A_{n+2,2;n+1,2}
-\left( \kappa (2n+1)+\gamma_f \right) A_{n+1,2;n,2} +\gamma A_{n+1,3;n,3}
\nonumber \\  &&+g\sqrt{n+1}A_{n,3;n,2}+g\sqrt{n}A_{n+1,2;n-1,3}\\
{{dA_ {n+3,1;n,3}} \over {dt}}&=&2\kappa \sqrt{(n+2)(n+1)}A_{n+2,3;n+1,3}
-\left( \kappa (2n+1)+\gamma_4 \right)A_{n+1,3;n,3}
\nonumber \\ &&-g\sqrt{n+2}A_{n+2,2;n,3}-g\sqrt{n+1}A_{n+1,3;n+1,2}\\
{{dA _{n+1,4;n,4}} \over {dt}}&=&2\kappa \sqrt{(n+2)(n+1)}A_{n+2,4;n+1,4}
-\left( \kappa (2n+1)+\gamma_f \right) A_{n+1,4;n,4}\nonumber \\
&&-E \left( A_{n+1,1;1,4} + A_{n+1,4;n,1} \right) \\
{{dA _{n+2,2;n,3}} \over {dt}}&=&2\kappa \sqrt{(n+3)(n+1)}A_{n+3,2;n+1,3}
-\left( \kappa (2n+2) +\gamma /2 +\gamma_f /2 \right) A {n+2,2;n,3}\nonumber 
\\
&&g\sqrt{n+2}A_{n+1,3;n,3}-g\sqrt{n+1}A_{n+2,2;n+1,2}\\
{{dA_ {n,3;n,2}} \over {dt}} &=&2\kappa (n+1)A_{n+1,3;n+1,2}-\left(
2\kappa n +\gamma /2 \gamma_f /2 \right) A_{n,3;n,2}\nonumber \\
&&+g\sqrt{n}A_{n,3;n-1,3}-g\sqrt{n+1}A_{n+1,2;n,3}\\
{{dA _{n+1,2;n,2}} \over {dt}}&=&2\kappa \sqrt{(n+2)(n+1)}A_{n+2,4;n+1,4}
-\left( \kappa (2n+1) +\gamma_4 \right)A_{n+1,2;n,2}\nonumber \\
&&+E \left( A_{n+1,4;n,4}-A_{n+1,1;n,1} \right).
\end{eqnarray}
As in the case of the incoherently pumped four-level laser,
\begin{equation}
\beta ={{2g^2/(\gamma_f+2\kappa )} \over {2g^2/(\gamma_f+2\kappa )+\gamma 
/2}}.
\end{equation}

Figure \ref{fig:sal19} presents the linewidth of the output spectrum for 
various values of $\beta$, with $\gamma_4 /\gamma=10.0, \kappa /\gamma=0.1$, 
and $\gamma_f /\gamma=10.0$ The results are in qualitative agreement with 
those of the incoherently pumped four-level laser, although there is a notable 
difference in the rate of initial increase/decrease. In Figure 
\ref{fig:sal20}, we plot the output spectrum in the regime where vacuum-Rabi 
structures are present. Again, the results are qualitatively the same as the 
incoherently pumped model, with the persistence of vacuum-Rabi structures for 
large pumps and mean intracavity photon numbers above unity.

\section{Conclusions}
\label{sec:sal5}
\ \indent In this chapter we have examined the output spectrum of several 
types of single atom laser systems. For the incoherently pumped three-level 
model, we find that for atom-field couplings at the lower range of that needed 
to produce photons in the cavity, that the spectrum is approximately a 
Lorentzian with a width broader than the Schwalow-Townes width. The width 
initially decreases with increasing pump strength as the photon number 
increases. The laser linewidth then expands with further increases in pump 
strength, and the photon number decreases with increasing pump strength. If 
the atom-field coupling is increased so that $\beta=0.5$, the laser emits 
amplitude squeezed light. Since the output field of a laser is not a minimum 
uncertainty state due to phase diffusion, it is not necessary that with 
decreases in amplitude noise the phase noise (linewidth) must increase, but it 
does so here. If we increase the atom field coupling to a value larger than 
all the other rates in the system, we find a vacuum-Rabi structure in the 
output structure as predicted by Loffler et. al. \cite{Loffler} This structure 
persists only for very small pump rates, as the incoherent pump rapidly 
decoheres the induced dipole. At moderate to larger pump rates, the spectrum 
is single-peaked.

The incoherently pumped four-level laser also has a single peaked spectrum 
for smaller atom-field couplings that is approximately Lorentzian. The 
linewidth decreases with the inverse of the photon number, but is always 
broader than the Schawlow-Townes limit. The photon and linewidth both pin at 
asymptotic values as the pump is increased. This is due to the fact that it 
does no good to pump the system at a rate faster than the ground state of the 
atom is replenished by decay from the lower lasing level. As $\beta$ is 
increased to $0.5$ or so, the system emits amplitude squeezed light, and the 
linewidth increases with pump strength as in the case of the three-level 
system. Finally, as $g$ is made larger than all the other rates in the system, 
the output spectrum has a vacuum-Rabi structure that persists for large pumps, 
even when the mean intracavity photon number is greater than unity. In this 
regime, for small pump rates, the spectrum is single-peaked however, even 
though $g$ is the largest rate, which might suggest a double-peaked spectrum. 
This has been explained using quantum trajectory simulations.

We have further considered a coherently pumped four-level single atom laser; 
the results are very similar to those of the incoherently pumped four-level 
laser. It is hoped that with recent advances in experimental techniques 
\cite{Kimble} that these types of systems will be examined in the laboratory 
soon.

\begin{figure}
\caption{Schematic diagram of single three-level atom in a cavity with 
incoherent pump, and level 4 adiabatically eliminated. For the 4-level system, 
$\gamma_{ij}$ 's are spontaneous emission rates from level $i $ to $j$,   
$\Gamma^\prime$ is a pump rate. For the 3-level system  $\Gamma $ is an 
effective pump rate, $\gamma $ is the spontaneous emission rate on the lasing 
transition. For both systems, $\kappa $ is the cavity decay rate and $g$ is 
the atom-field coupling strength.}
\label{fig:sal1}
\end{figure}

\begin{figure}
\caption{The output spectrum of the single three-level incoherently pumped 
laser, for $g/\gamma=\kappa/\gamma=0.1$ as a function of pumping strength 
$\Gamma /\gamma$.}
\label{fig:sal2}
\end{figure}

\begin{figure}
\caption{The linewidth of the single three-level incoherently pumped laser, as 
a function of pumping strength $\Gamma /\gamma $ for $\kappa/\gamma=0.1$ and 
$g/\gamma=0.6$ (solid line). The dashed line is a plot of $\kappa /2\langle n 
\rangle $ for the same parameters.}
\label{fig:sal3}
\end{figure}

\begin{figure}
\caption{Mean intracavity photon number for the incoherently pumped 
three-level laser, as a function of pumping strength $\Gamma /\gamma$, for  
$\kappa/\gamma=0.1$ and $g/\gamma=0.6$.}
\label{fig:sal4}
\end{figure}

\begin{figure}
\caption{The output spectrum of the single three-level incoherently pumped 
laser, for $g/\gamma=1.414$ and $\kappa/\gamma=0.1$ as a function of pumping 
strength $\Gamma /\gamma$. This plot is for small pumping strengths.}
\label{fig:sal5}
\end{figure}

\begin{figure}
\caption{The output spectrum of the single three-level incoherently pumped 
laser, for $g/\gamma=1.414$ and $\kappa/\gamma=0.1$ as a function of pumping 
strength $\Gamma /\gamma$. Here we show the behavior over a broad range of 
pump values.}
\label{fig:sal6}
\end{figure}

\begin{figure}
\caption{Mean intracavity photon number for the incoherently pumped 
three-level laser, as a function of pumping strength $\Gamma /\gamma$, for  
$\kappa/\gamma=0.1$ and $g/\gamma=1.414$.}
\label{fig:sal7}
\end{figure}

\begin{figure}
\caption{(a) Plot of the conditioned population of the upper level of the 
lasing transition for the three-level incoherently pumped laser with $\Gamma 
/\gamma =1.0$, $g/\gamma=1.414$, and $\kappa /\gamma=0.1$\\
(b) Plot of the conditioned induced dipole on the lasing transition the 
three-level incoherently pumped laser with $\Gamma /\gamma =1.0$, 
$g/\gamma=1.414$, and $\kappa /\gamma=0.1$.}
\label{fig:sal8}
\end{figure}

\begin{figure}
\caption{(a) Plot of the conditioned population of the upper level of the 
lasing transition for $\Gamma /\gamma =10.0$, $g/\gamma=0.6$, and $\kappa 
/\gamma=0.1$.\\
(b) Plot of the conditioned induced dipole on the lasing transition for 
$\Gamma /\gamma =10.0$, $g/\gamma=0.6$, and $\kappa /\gamma=0.1$.}
\label{fig:sal9}
\end{figure}

\begin{figure}
\caption{ Schematic diagram of single four-level atom in a cavity with 
incoherent pump, and level 4 adiabatically eliminated. For the 4-level system, 
$\gamma_{ij}$ 's are spontaneous emission rates from level $i $ to $j$,   
$\Gamma^\prime$ is a pump rate. For the 3-level system  $\Gamma$ is an 
effective pump rate, $\gamma$ is the spontaneous emission rate on the lasing 
transition, and  $\gamma_f $ is the spontaneous emission rate from the lower 
lasing level. For both systems, $\kappa$ is the cavity decay rate and $g$ is 
the atom-field coupling strength.}
\label{fig:sal10}
\end{figure}

\begin{figure}
\caption{The output spectrum of the single four-level incoherently pumped 
laser, for $g/\gamma=\kappa/\gamma=0.1$, and $\gamma_f /\gamma=10.0$ as a 
function of pumping strength $\Gamma /\gamma$.}
\label{fig:sal11}
\end{figure}

\begin{figure}
\caption{The linewidth of the single four-level incoherently pumped laser, as 
a function of pumping strength $\Gamma /\gamma $ for $\kappa/\gamma=0.1$ and 
$g/\gamma=0.6$ for (a) $\beta=0.3$, (b) $\beta =0.4$, and (c)$\beta=0.5$. The 
dashed lines are plots of $\kappa /2\langle n \rangle $ for the same 
$\gamma_f /\gamma, \kappa /\gamma$ and (d) $\beta=0.3$, (e) $\beta =0.4$, and 
(f)$\beta=0.5$.}
\label{fig:sal12}
\end{figure}

\begin{figure}
\caption{The linewidth of the single four-level incoherently pumped laser, as 
a function of pumping strength $\Gamma /\gamma $ for $\kappa/\gamma=0.1$ and 
$g/\gamma=0.6$ for (a) $\beta=0.6$, (b) $\beta =0.7$, (c) $\beta=0.8$,  and 
(d) $\beta=0.9$. In (e) $\beta=0.998$, with $\gamma_f /\gamma=100.0$ and 
$\kappa /\gamma=0.1$.}
\label{fig:sal13}
\end{figure}

\begin{figure}
\caption{The output spectrum of the single four-level incoherently pumped 
laser, for $g/\gamma=10.0$, $\kappa/\gamma=0.1$, and $\gamma_f /\gamma=2.0$ as 
a function of pumping strength $\Gamma /\gamma$.}
\label{fig:sal14}
\end{figure}

\begin{figure}
\caption{Mean intracavity photon number for the incoherently pumped four-level 
laser, as a function of pumping strength $\Gamma /\gamma$, for  
$\kappa/\gamma=0.1$, $\gamma_f /\gamma=2.0$, and $g/\gamma=10.0$.}
\label{fig:sal15}
\end{figure}

\begin{figure}
\caption{(a) Plot of the conditioned population of the upper level of the 
lasing transition the four-level incoherently pumped laser with $\Gamma 
/\gamma =1.0$, $g/\gamma=10.0$, $\gamma_f /\gamma=2.0$, and $\kappa 
/\gamma=0.1$.\\
(b) Plot of the conditioned induced dipole on the lasing transition for 
$\Gamma /\gamma =1.0$, $g/\gamma=10.0$, and $\kappa /\gamma=0.1$.}
\label{fig:sal16}
\end{figure}

\begin{figure}
\caption{(a) Plot of the conditioned population of the upper level of the 
lasing transition the four-level incoherently pumped laser with $\Gamma 
/\gamma =10.0$, $g/\gamma=10.0$, $\gamma_f /\gamma=2.0$, and $\kappa 
/\gamma=0.1$.\\
(b) Plot of the conditioned induced dipole on the lasing transition for 
$\Gamma /\gamma =10.0$, $g/\gamma=10.0$, and $\kappa /\gamma=0.1$.}
\label{fig:sal17}
\end{figure}

\begin{figure}
\caption{Schematic diagram of single four-level atom in a cavity with a 
coherent pump.  $E$ is an effective pump rate, $\gamma$ is the spontaneous 
emission rate on the lasing transition, $\gamma_4$ is the spontaneous emission 
rate out of the upper pumping level, and  $\gamma_f $is the spontaneous 
emission rate from the lower lasing level. Also, $\kappa$ is the cavity decay 
rate and $g$ is the atom-field coupling strength.}
\label{fig:sal18}
\end{figure}

\begin{figure}
\caption{The linewidth of the singlefour-level coherently pumped laser, as a 
function of pumping strength $\Gamma /\gamma $ for $\kappa/\gamma=0.1$ and 
$g/\gamma=0.6$ for (a) $\beta=0.6$, (b) $\beta =0.7$, (c) $\beta=0.8$,  and 
(d) $\beta=0.9$. In (e) $\beta=0.998$, with $\gamma_f /\gamma=100.0$ and 
$\kappa /\gamma=0.1$.}
\label{fig:sal19}
\end{figure}

\begin{figure}
\caption{The output spectrum of the single four-level coherently pumped laser, 
for $g/\gamma=10.0$, $\kappa/\gamma=0.1$, and $\gamma_f /\gamma=2.0$ as a 
function of pumping strength $\Gamma /\gamma$.}
\label{fig:sal20}
\end{figure}

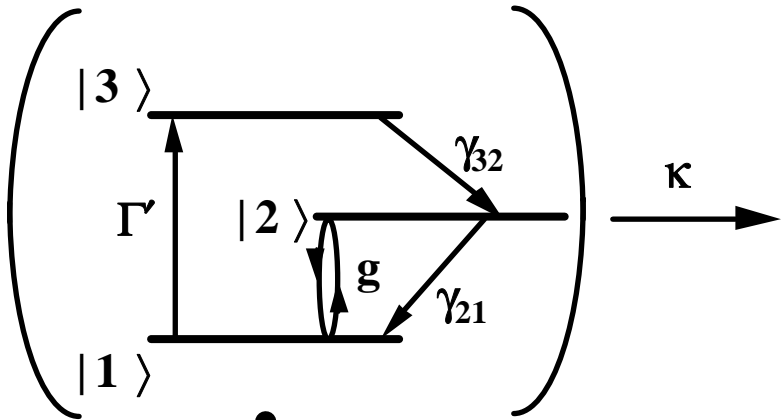
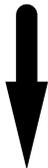
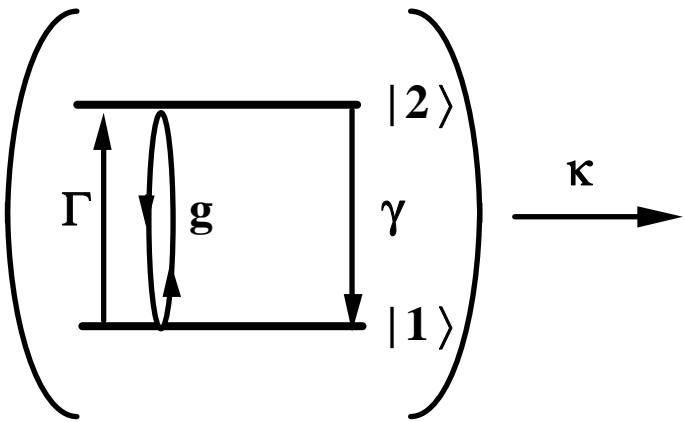
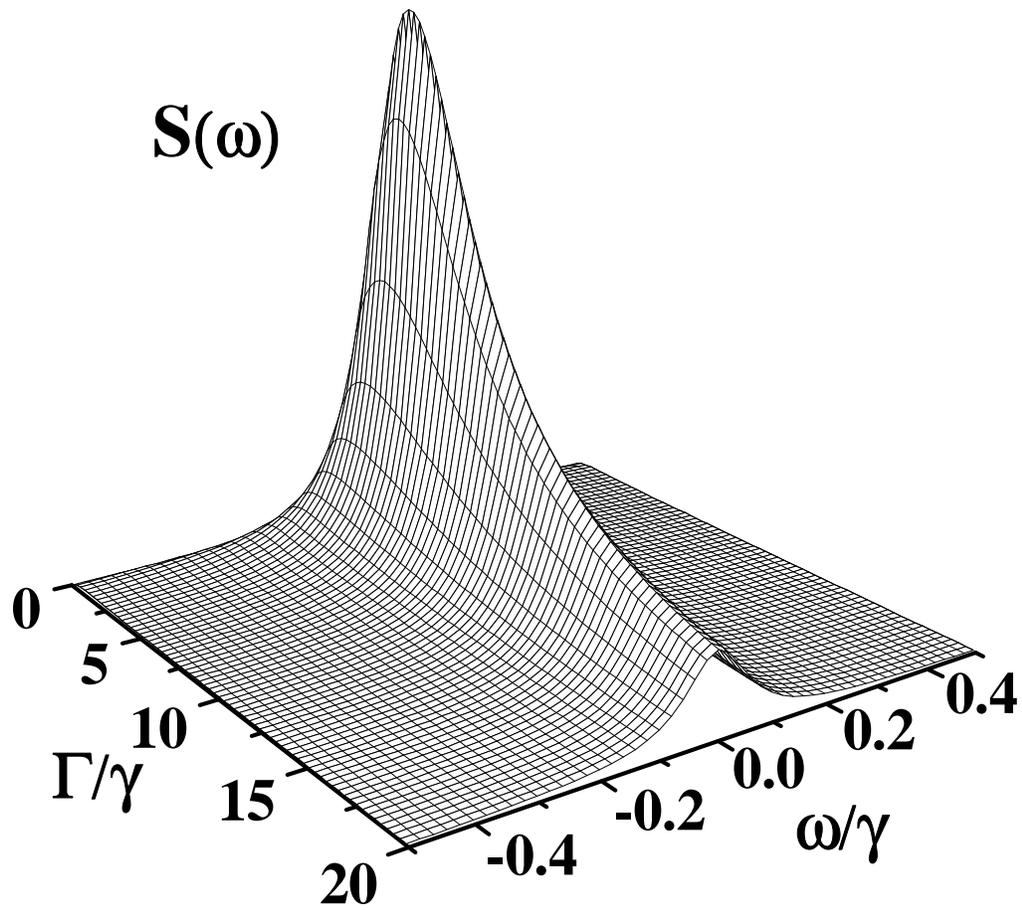
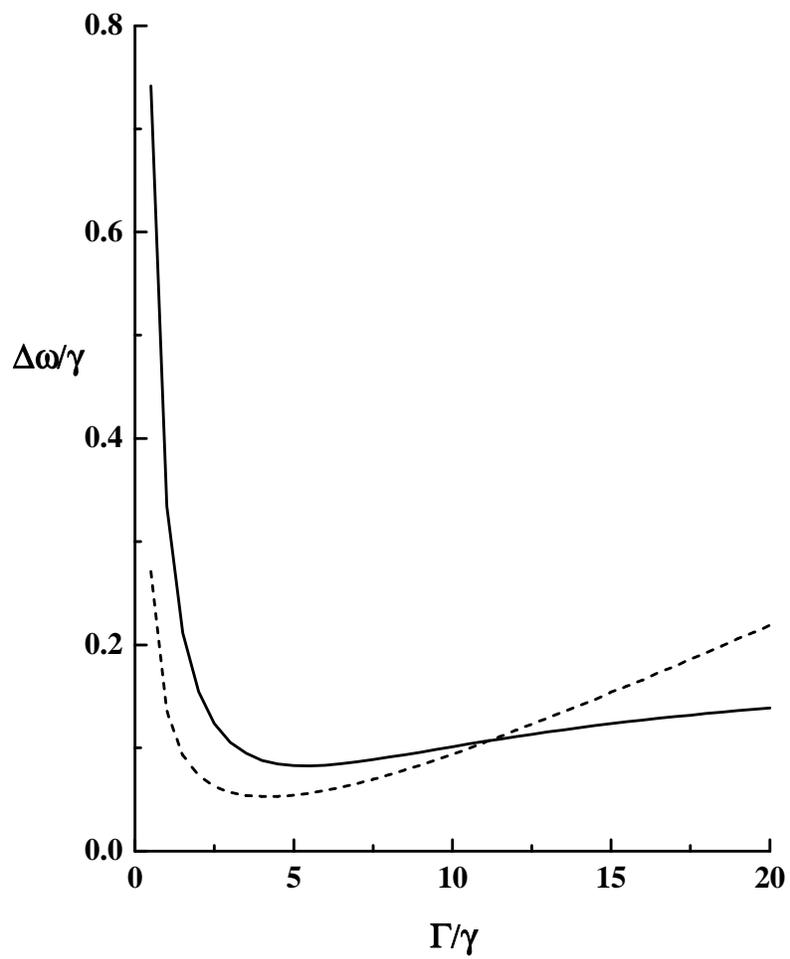
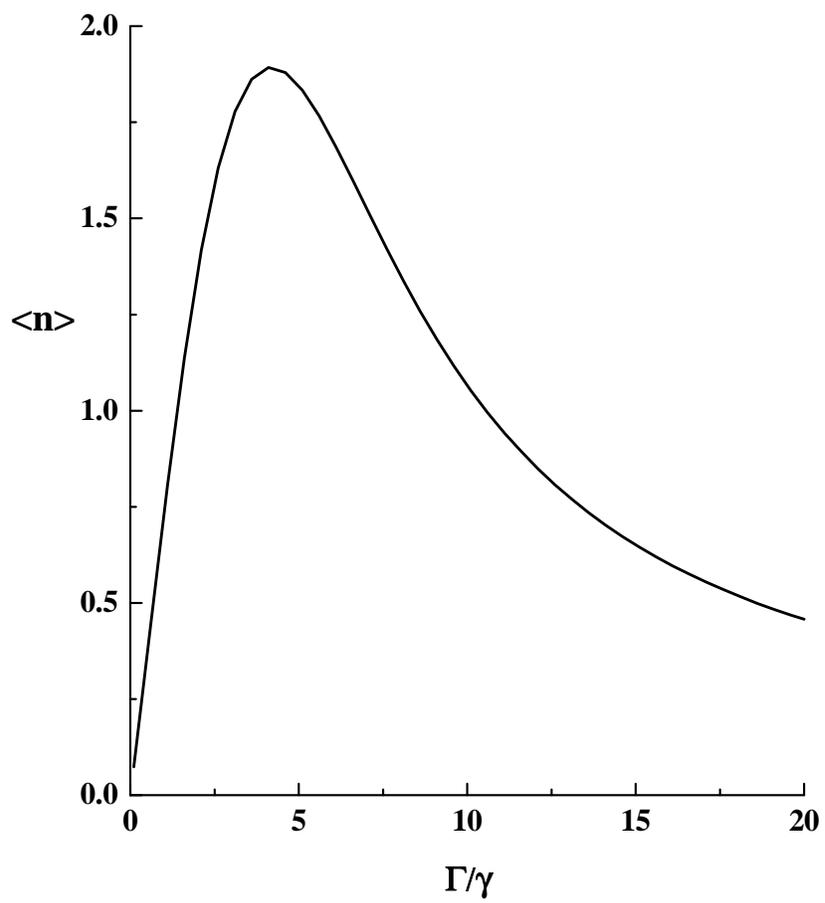
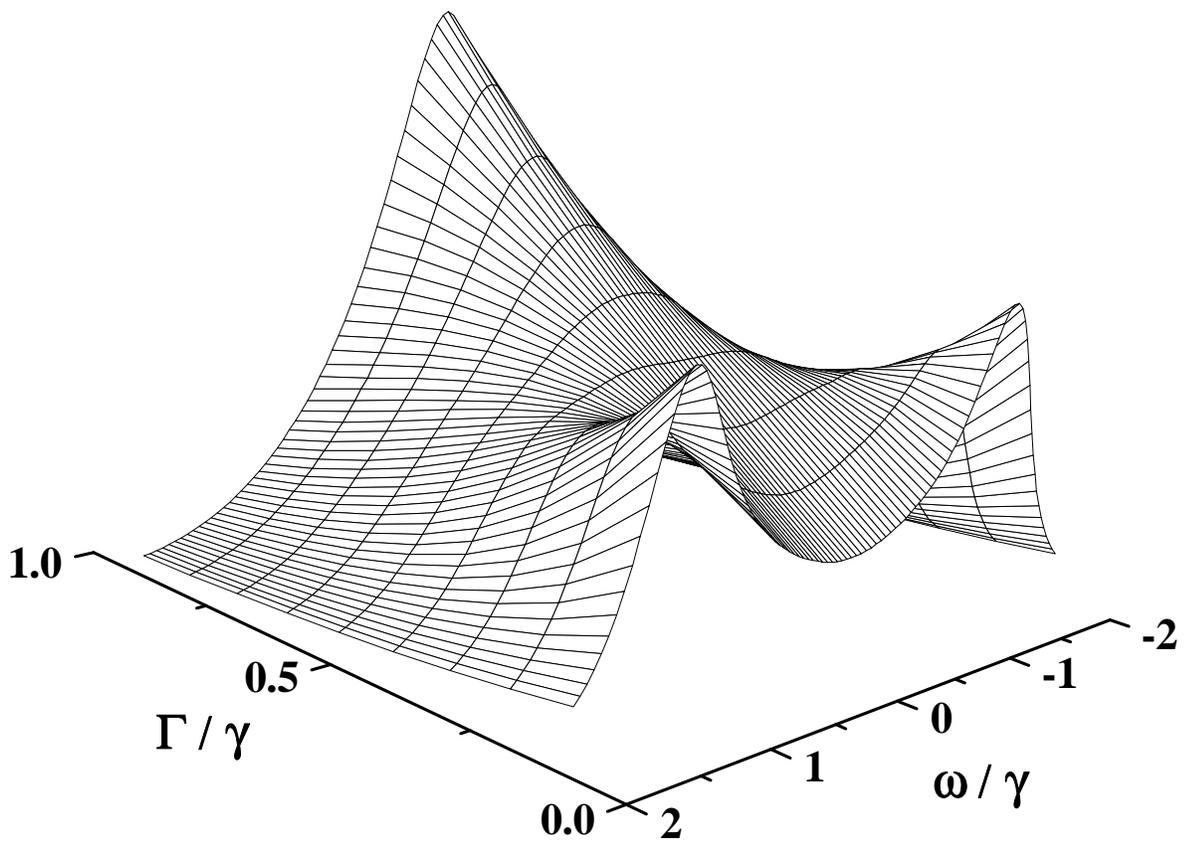
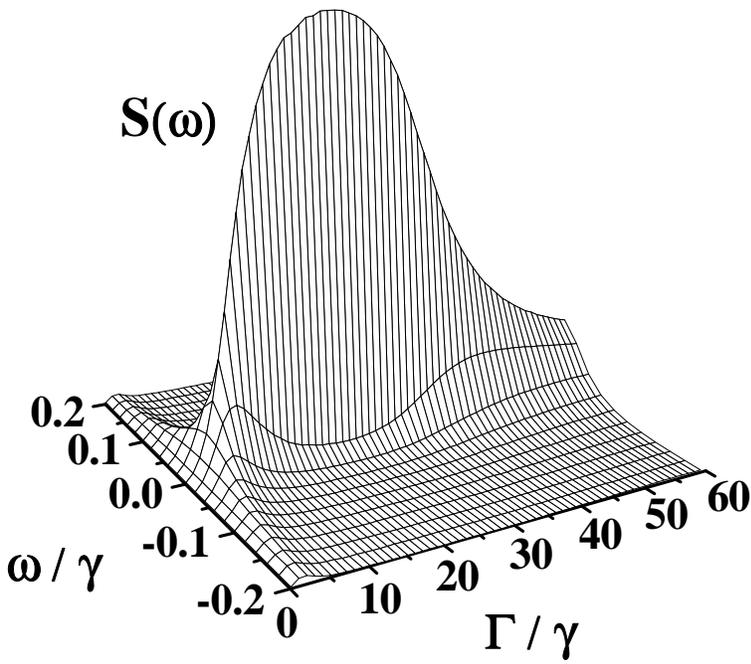
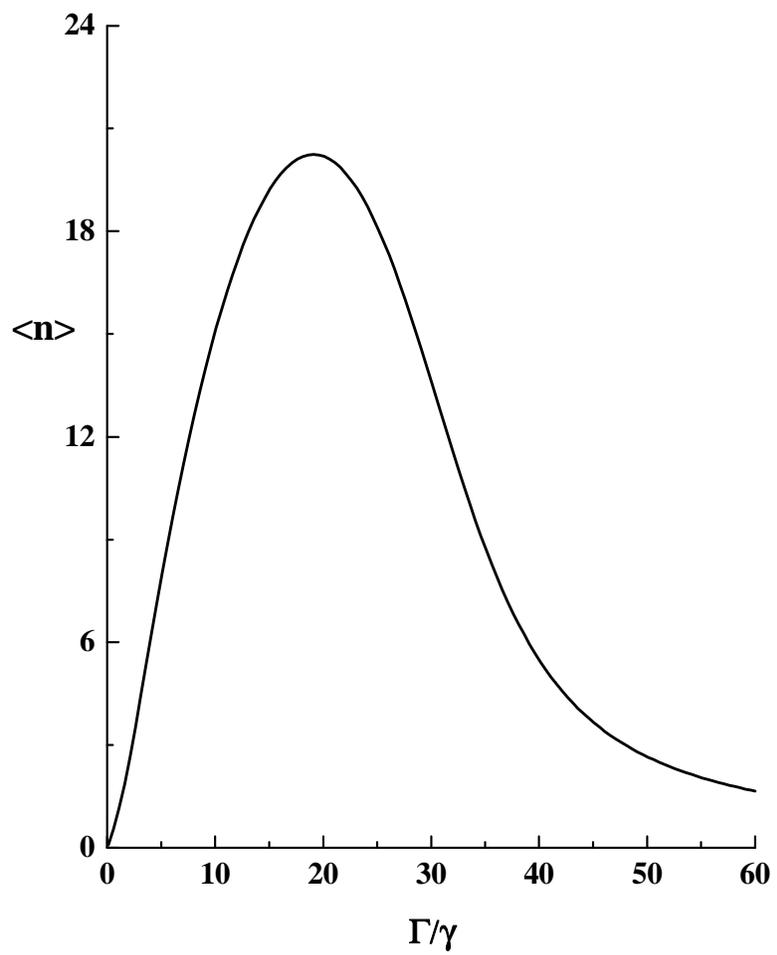
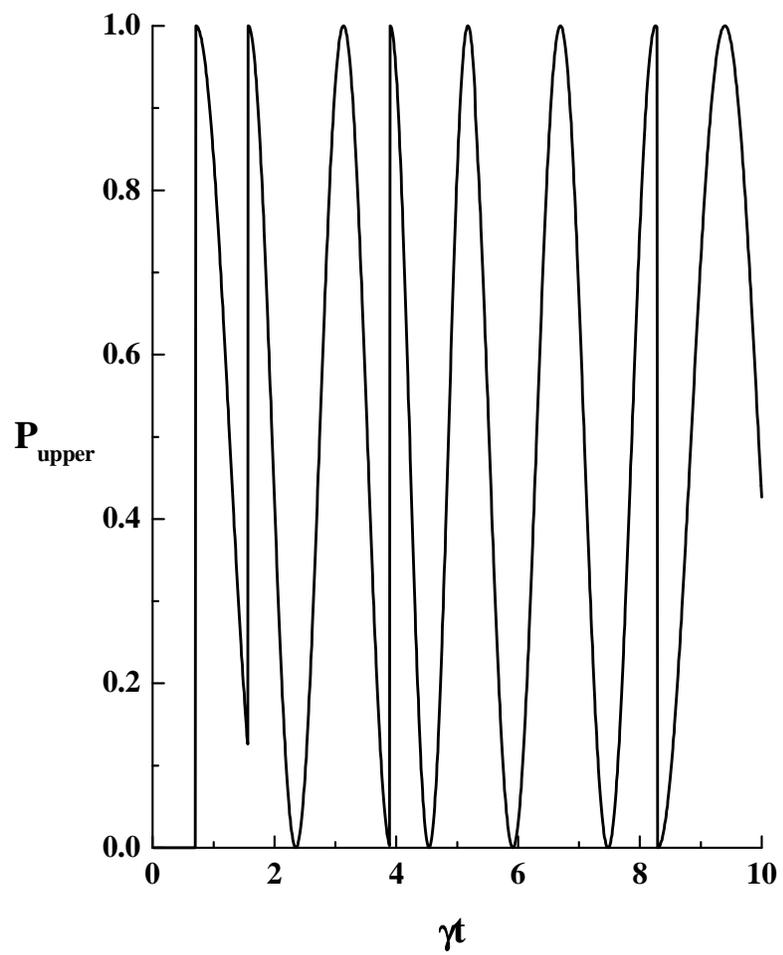
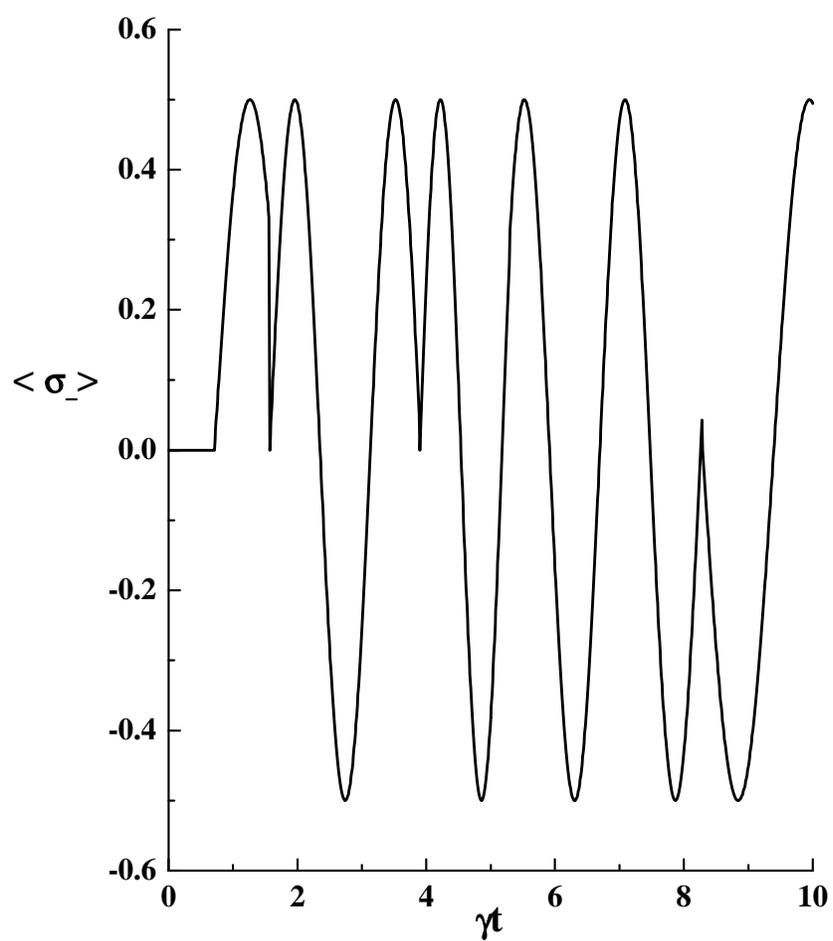
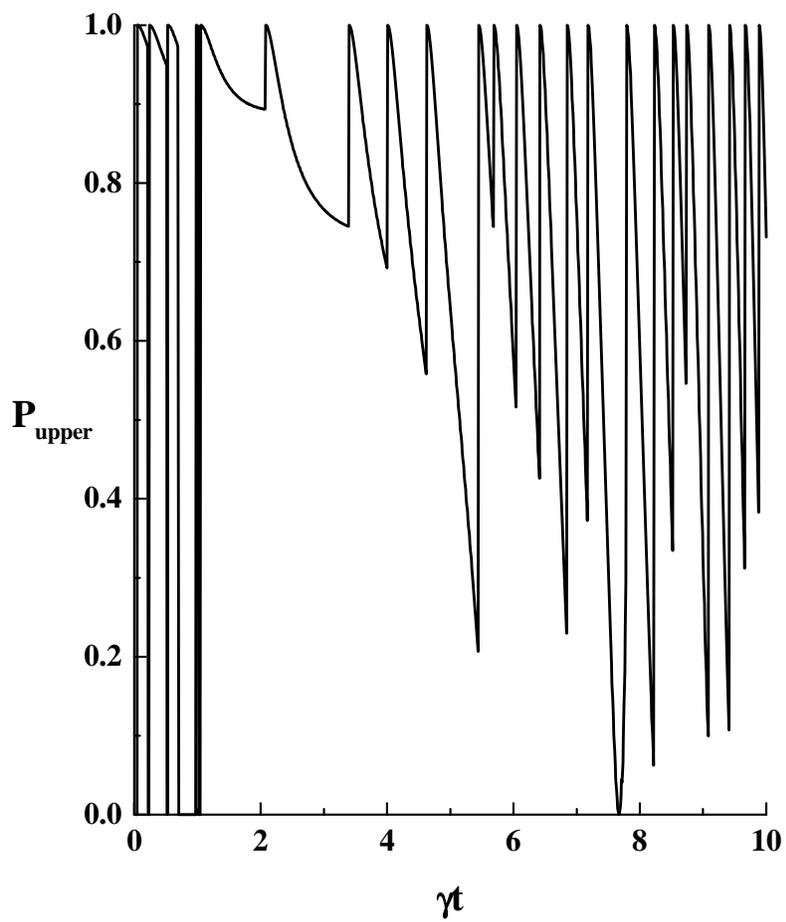
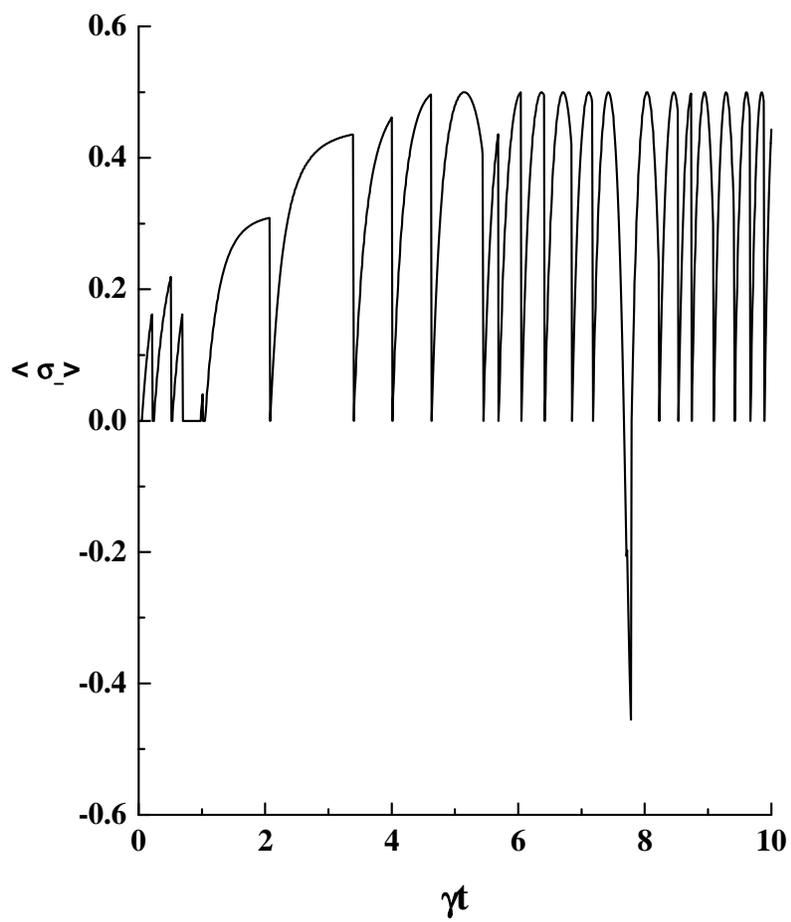
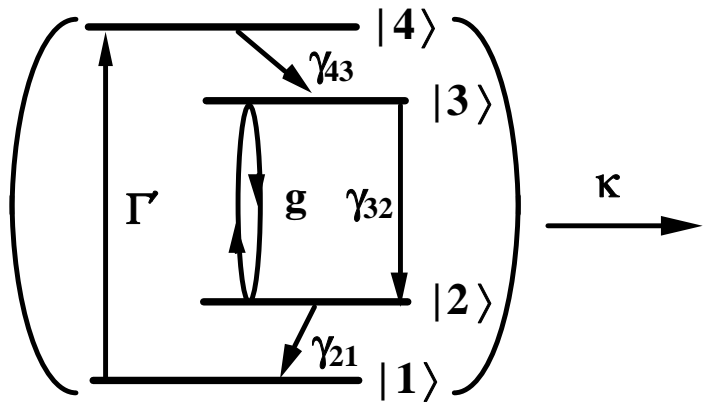
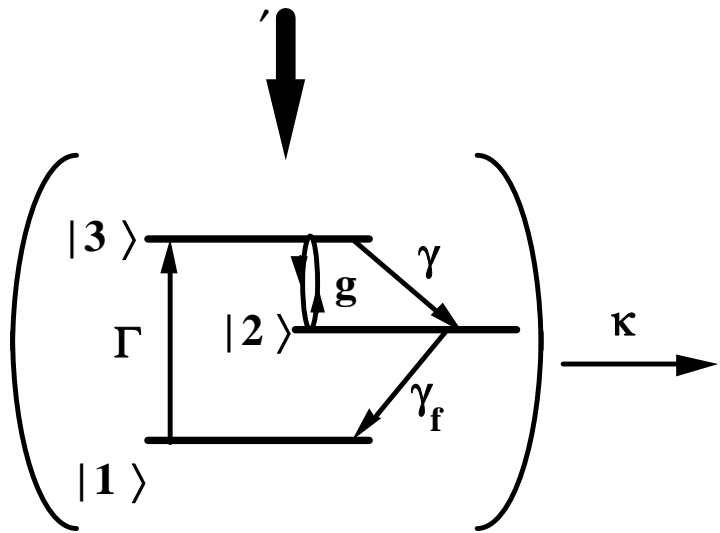
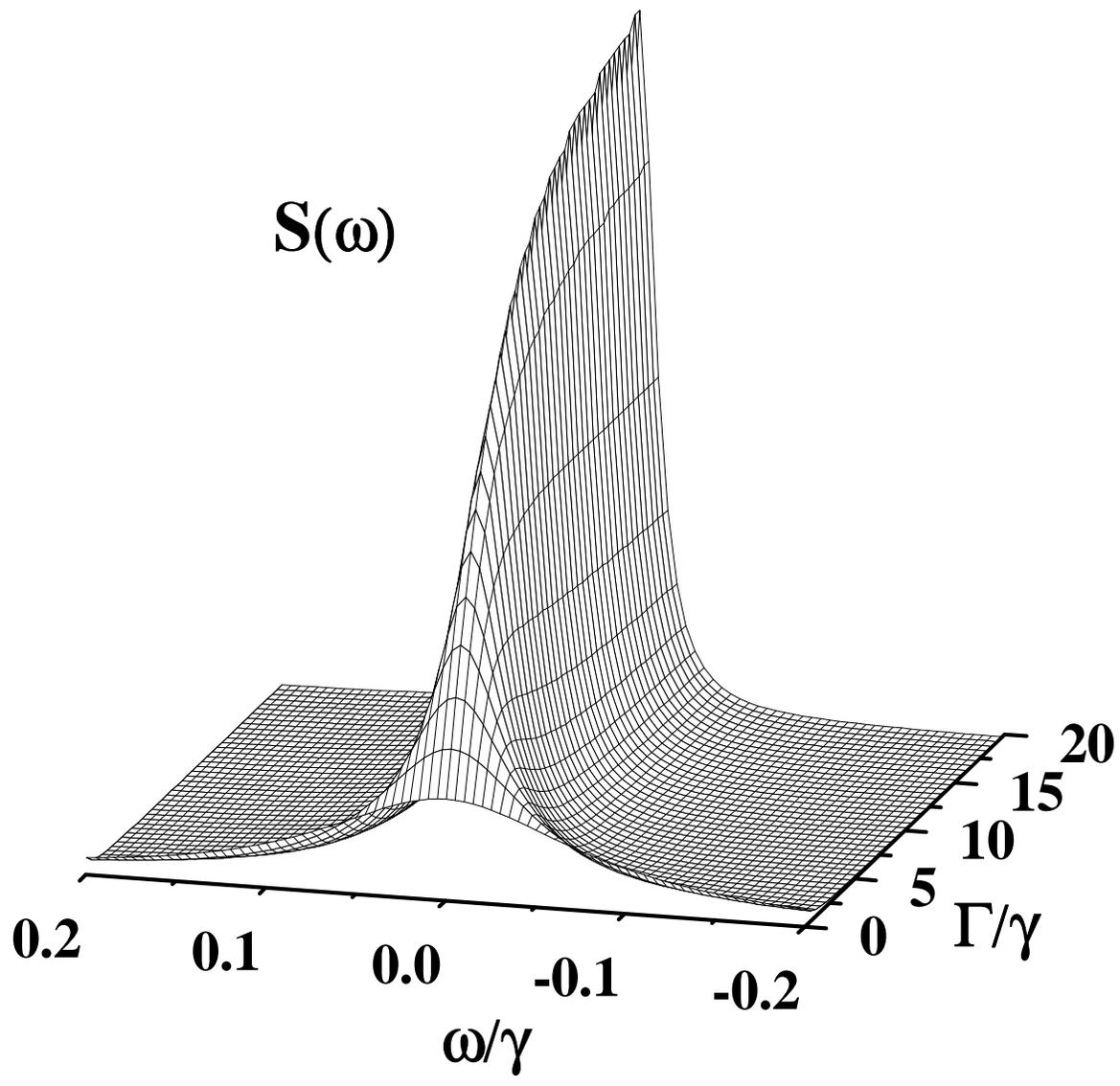
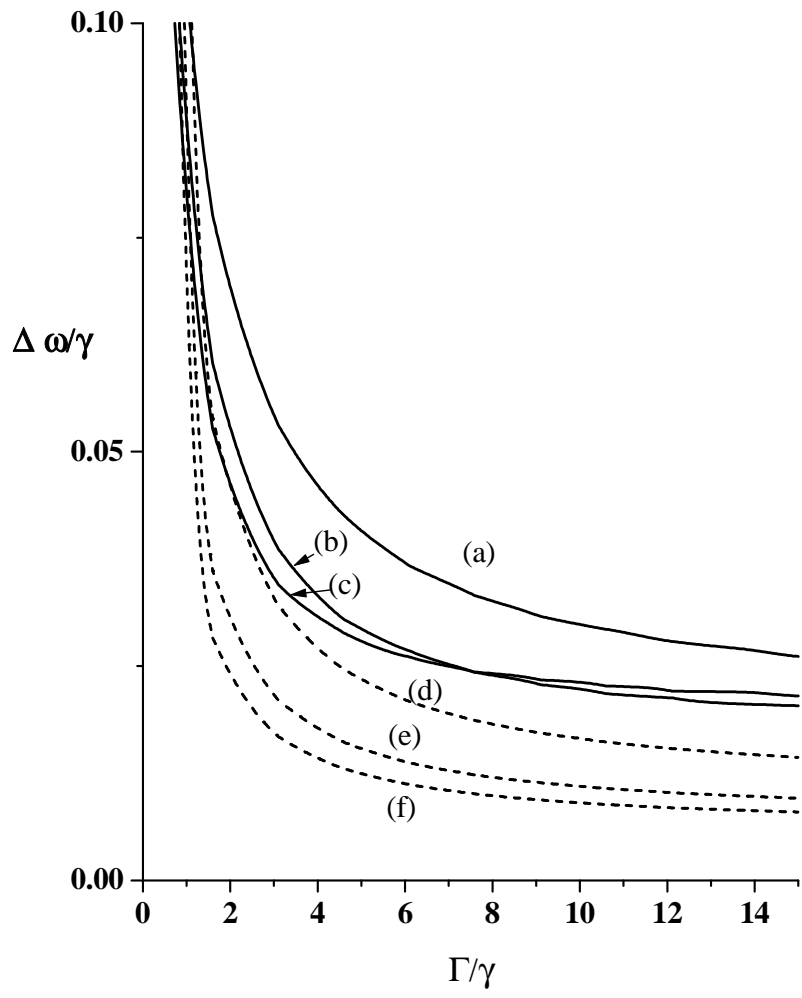
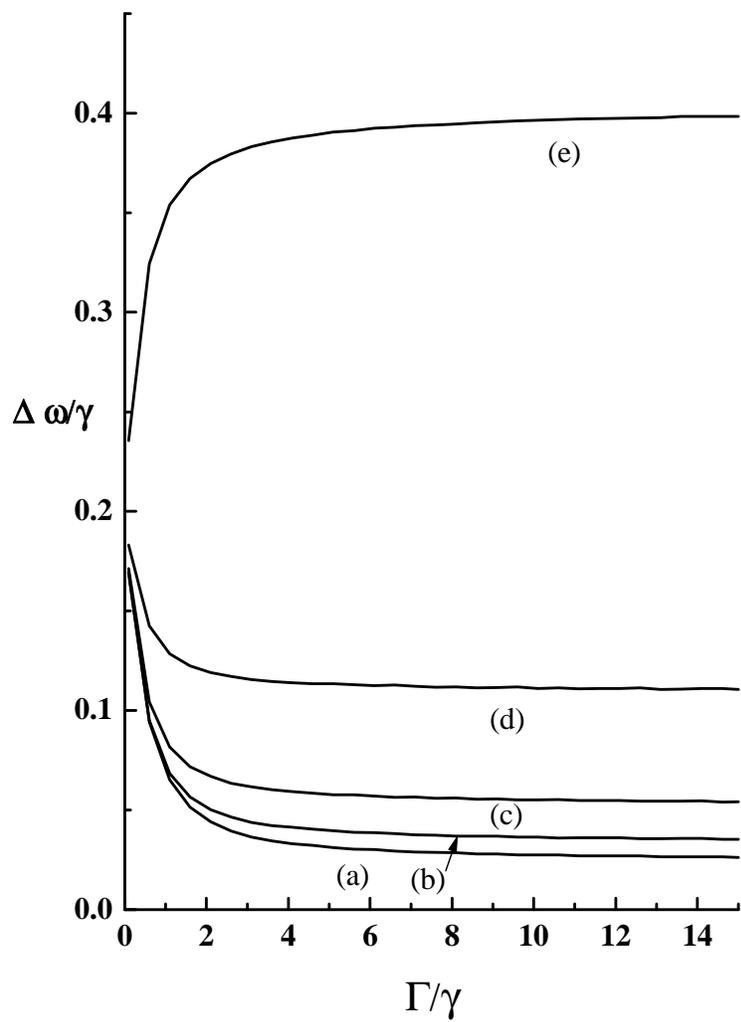
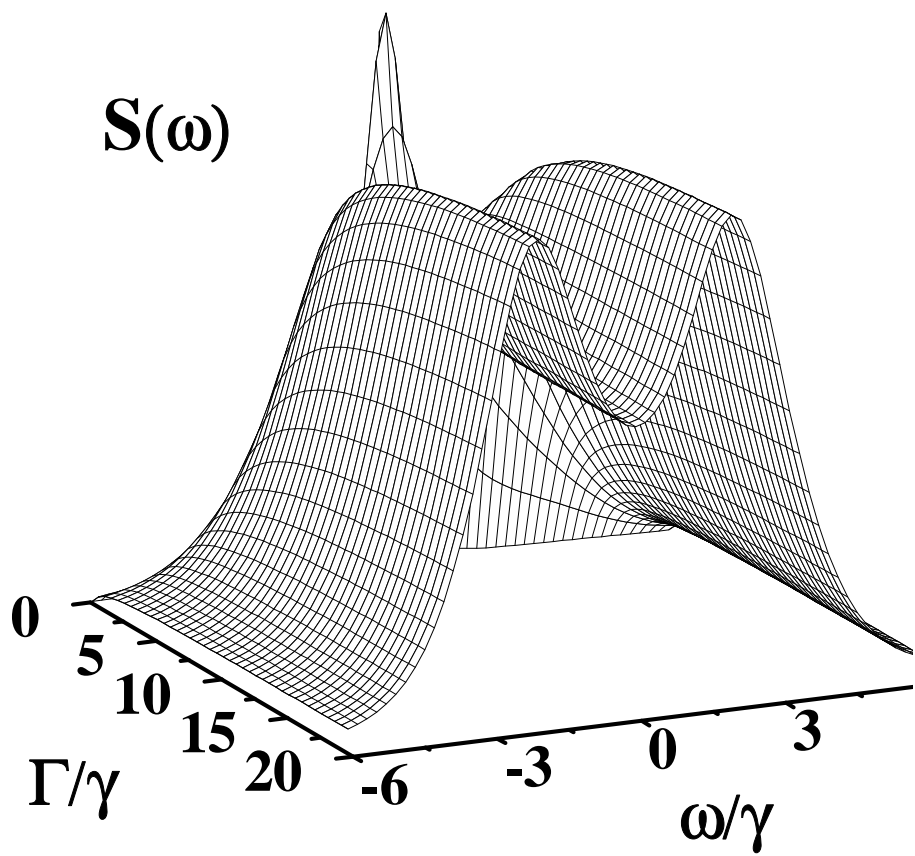
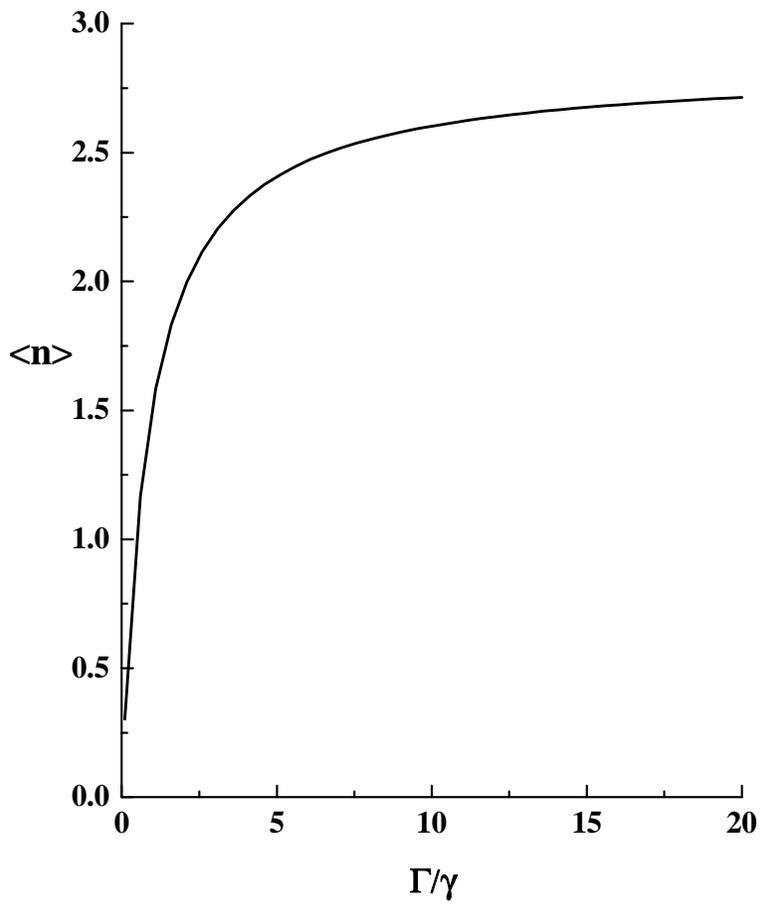
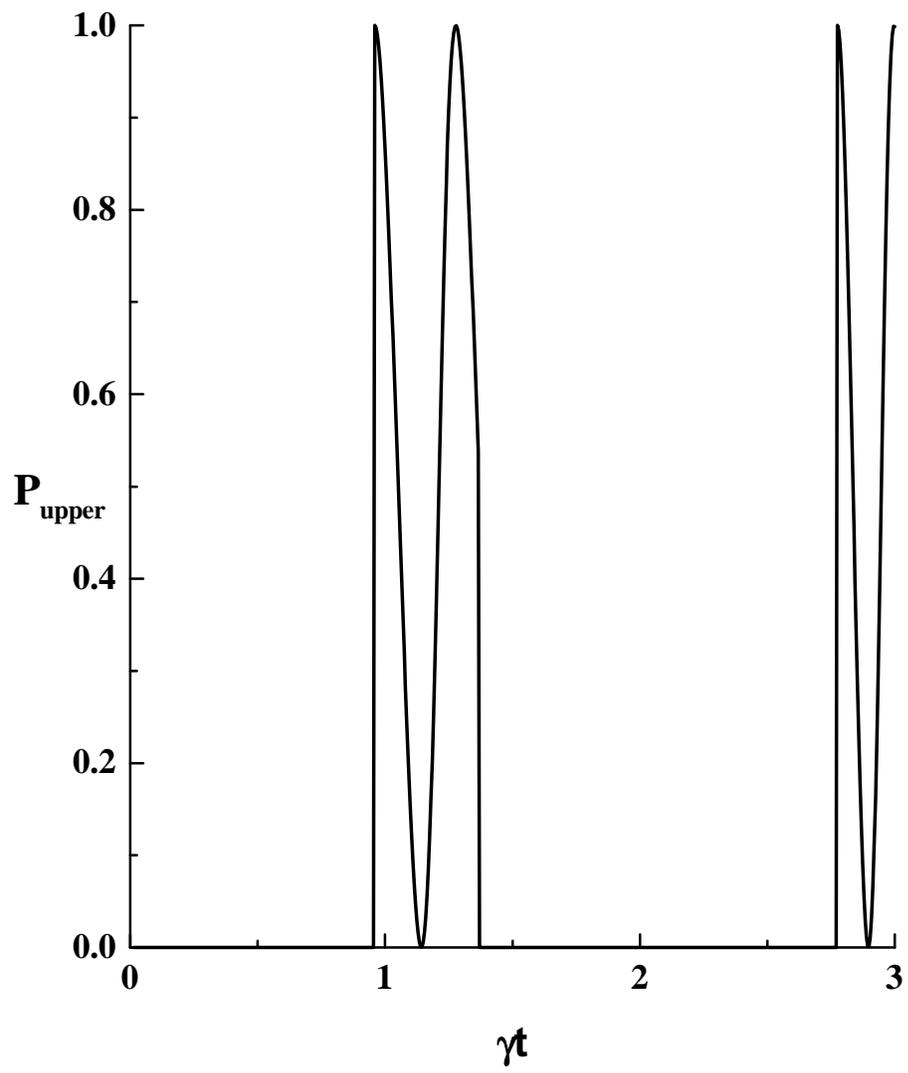
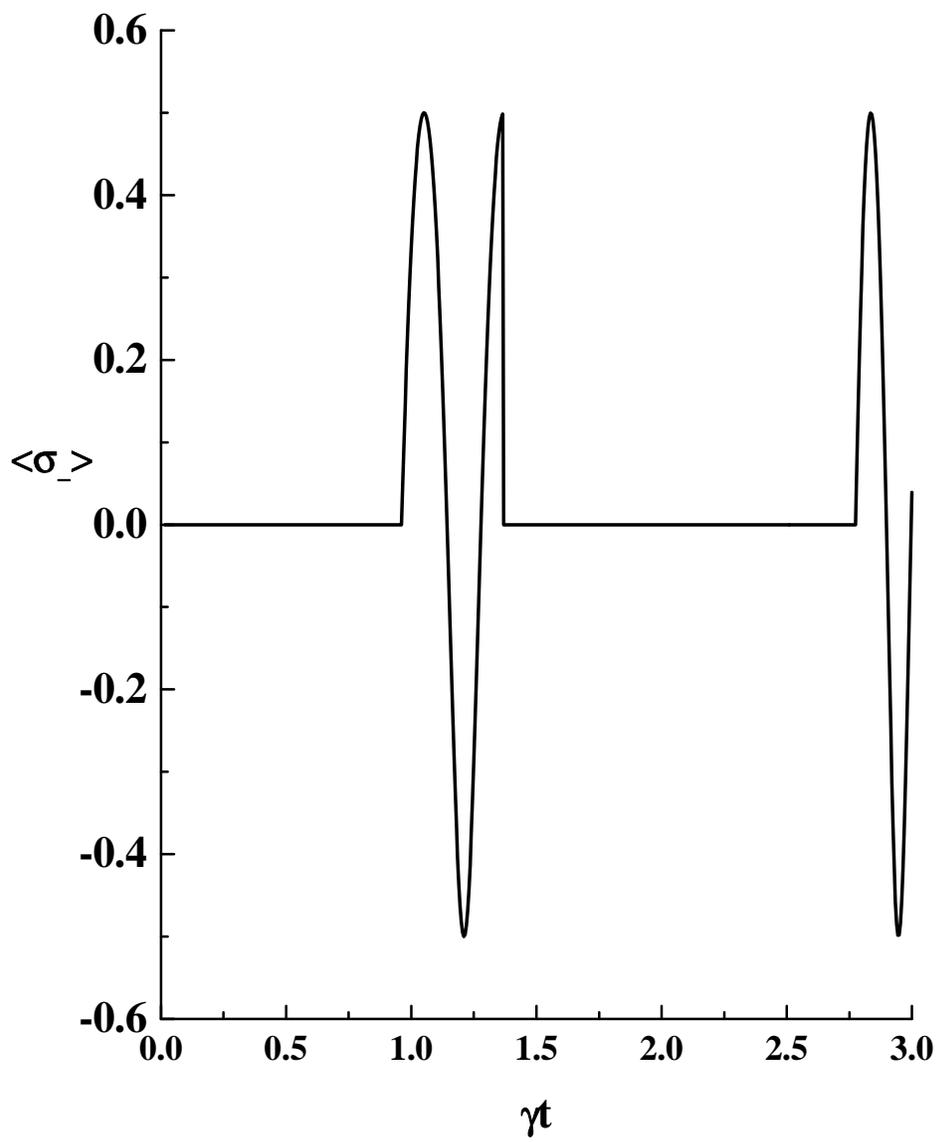
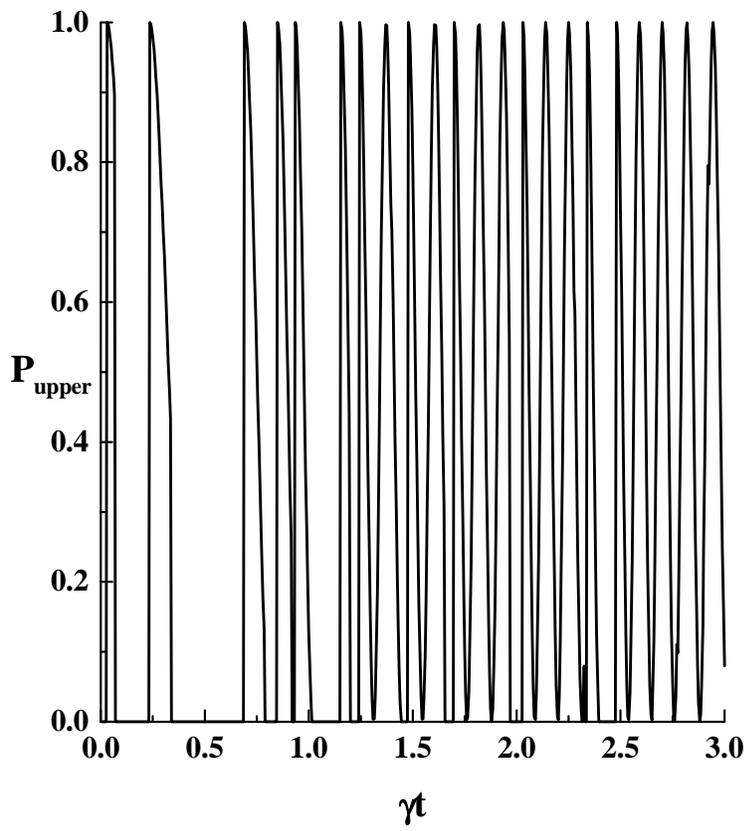
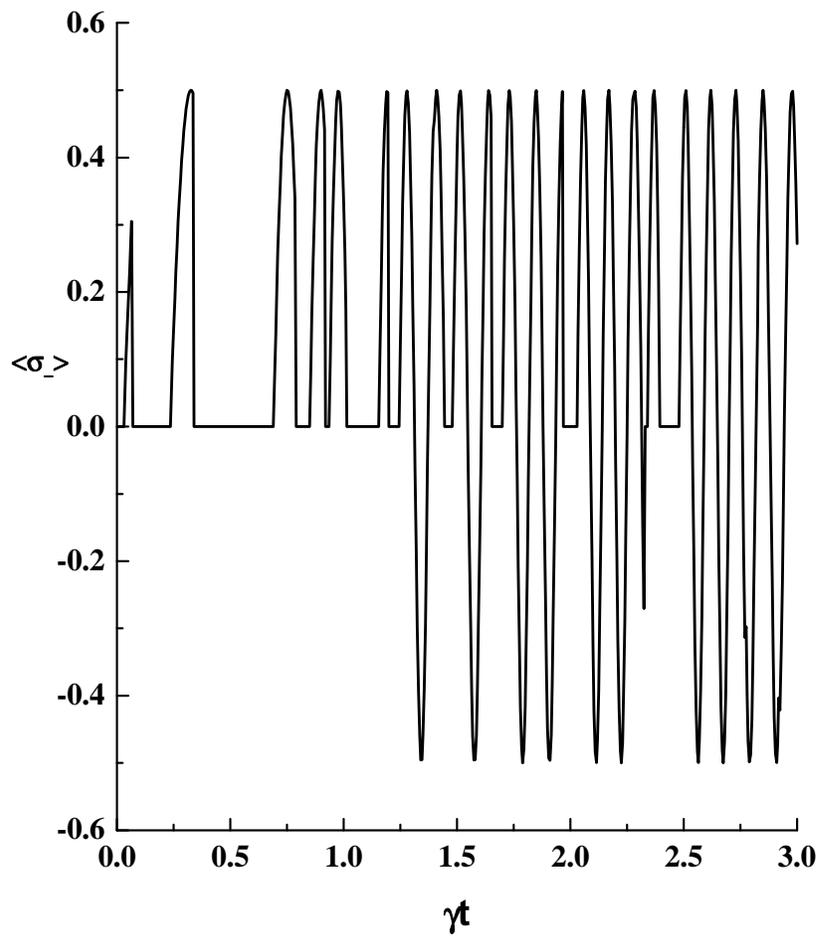
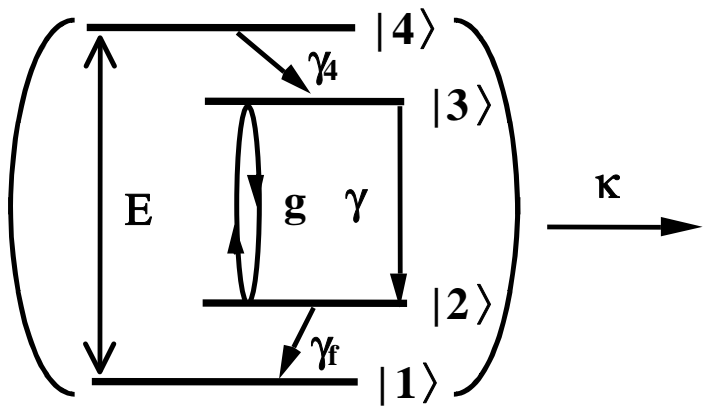
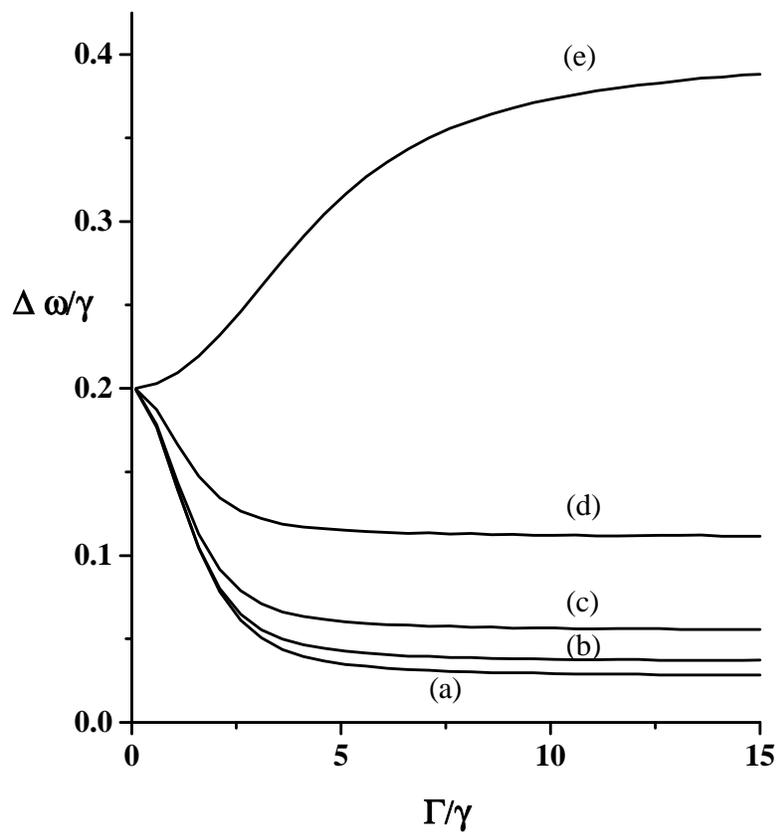
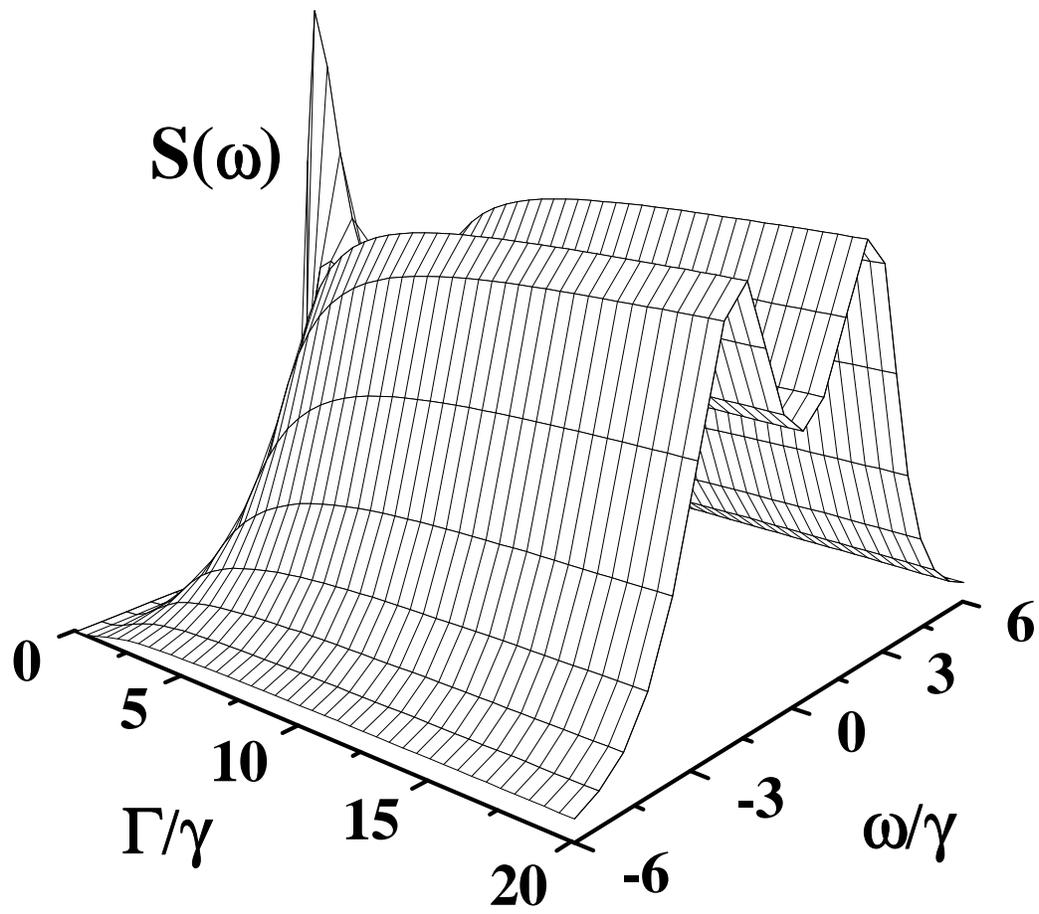
\end{document}